\documentclass[prd,twocolumn,aps,longbibliography,notitlepage,superscriptaddress]{revtex4}
\usepackage{amsfonts}
\usepackage{mathrsfs}
\usepackage{amsmath}
\usepackage{color}
\usepackage{graphicx}
\usepackage{bm}
\usepackage{amssymb}
\usepackage{xspace}
\usepackage{epstopdf}
\usepackage{dcolumn}
\usepackage{longtable}
\usepackage{multirow}
\usepackage{ulem}
\usepackage{booktabs}
\usepackage[colorlinks=true, letterpaper=true, pdfstartview=FitV, linkcolor=blue, citecolor=blue, urlcolor=blue]{hyperref}

\newcommand{\bs}{\mathbf{s}}

\def\bs#1{\boldsymbol{#1}}

\begin{document}
\title{Topological domain-wall states from Umklapp scattering in twisted bilayer graphene}

\author {Juncheng Li}
\author{Cong Chen}
\email{cchen93@hku.hk}
\author {Wang Yao}
\email{wangyao@hku.hk}
\affiliation{New Cornerstone Science Laboratory, Department of Physics, University of Hong Kong, Hong Kong, China}
\affiliation{HK Institute of Quantum Science and Technology, University of Hong Kong, Hong Kong, China}

\begin{abstract}
Twistronics, harnessing interlayer rotation to tailor electronic states in van der Waals materials, has predominantly focused on small-angle regime. 
Here, we unveil the pivotal role of intervalley Umklapp scattering in large-angle twisted bilayer graphene, which governs low-energy physics and drives unconventional band topology. 
By constructing symmetry-constrained effective $k\cdot p$ models for $\pm 21.8^{\circ}$-twisted bilayers, we demonstrate how structural chirality imprints distinct electronic responses. The $D_6$ configuration exhibits a gapped spectrum with chiral interlayer coupling, while $D_3$ symmetric stacking configuration displays semimetallic behavior. 
Crucially, chirality inversion creates topological domain-wall states,  which manifest as counterpropagating pseudospin modes at interfaces between oppositely twisted regions.
These states, absent in untwisted bilayers, emerge from a Jackiw-Rebbi-like mechanism tied to chirality reversal. 
Atomistic simulations confirm these topological states and demonstrate their robustness against symmetry-breaking perturbations.
The interplay between twist-induced chirality and topology opens new pathways for engineering domain-wall states in twisted materials.
\end{abstract}

\maketitle

\section{Introduction}

In recent years, the introduction of moir\'{e}  patterns created by interlayer rotation or lattice mismatch in vertically stacked two-dimensional (2D) crystals has been extensively explored for manipulating electron behavior~\cite{MacD_2020_NatM,Rubio2021moire,Andrea2021marvels,Wilson2021excitons,Zhangfan_2022Nat}. 
The twisted-moir\'{e}-driven paradigm, now termed twistronics, enables continuous tuning of electronic structures through a single rotational degree of freedom~\cite{Twist_PRB_2017,Twistronics_2021,LuY2024ChemRev}.
To date, most investigations on twisted van der Waals (vdW) materials have focused on the small-angle twist regime~\cite{CaoY2018_SuperCon,CaoY_2018_Correlated,Umk_Nat_WY_2019,WuF_TMD_PRL_2019,wangCorrelatedElectronicPhases2020,caiSignaturesFractionalQuantum2023}. 
Taking twisted bilayer graphene (tBG) as an example, a slight interlayer rotation induces a small momentum-space separation between the Dirac cones of adjacent layers. This results in low-energy physics governed primarily by intravalley scattering of Dirac electrons between neighboring layers~\cite{MacD_PNAS_2011,MacD_2020_NatM}.
Increasing the twist angle $\theta$ produces two distinct momentum-space effects: First, Dirac points from the same valley in adjacent layers get widely separated, restricting the intravalley interlayer hybridization to regions far from the Dirac point where Dirac cones overlap. Second, opposite valleys from adjacent layers approach each other beyond the first Brillouin zone, enabling the moir\'{e}-lattice-assisted interlayer tunneling through Umklapp processes~\cite{MacD_2010_PRB,Umk_PRL_2015,Umk_Nat_2019,Umk_Nat_WY_2019,Umk_NP_2019}.
This mechanism, critical for hybridizing opposite valleys from different layers, governs emergent topological phases such as 2D second-order topological insulator phase~\cite{Lee_tBGHOTI_2019,WangZF_2021_PRL} and three-dimensional (3D) chiral topological semimetal phase~\cite{CCLiu_2023PRB,Chen_2025_ROPP}, has yet to be fully deciphered.
 
Here, we systematically investigate the low-energy physics of large-angle tBGs ($\theta=21.8^{\circ}$), where \textit{intervalley} Umklapp scattering dominates. 
By constructing symmetry-adapted $k\cdot p$ models for different stackings belonging to chiral point groups $D_6$ or $D_3$, we reveal how structural chirality ($\zeta=\pm$) splits otherwise degenerate states and dictates band topology.
For example, the $D_6$-symmetric configuration exhibits a gapped spectrum with hidden Bernevig-Hughes-Zhang-like topology, while the $D_3$ stacking permits semimetallic phases with quadratic band crossings.
Furthermore, we demonstrate that chirality reversal across domain walls  generates topological protected interfacial states. These states emerge from a Jackiw-Rebbi-like mechanism, where the sign reversal in structural chirality ($\zeta$) creates a valley-projected pseudospin texture that supports counterpropagating modes. 
Numerical tight-binding calculations confirm their persistence against different symmetry-breaking perturbations and also demonstrate their robustness across most tBG configurations at $\theta=21.8^{\circ}$.
Our work unveils a novel approach to achieve robust topological channels at the domain-wall boundary, with implications for chiral transport engineering in twisted vdW materials.

The rest of the paper is organized as follows.  
In Section~\ref{structure}, we characterize the crystal structures of commensurate tBGs at $\theta=21.8^\circ$  with the symmetry of point group $D_6$ and $D_3$.  
Section~\ref{sec:effective_low_energy_models} builds on this foundation to develop effective low-energy $k \cdot p$ models for those configurations, uniquely incorporating the \textit{intervalley} Umklapp tunneling, capturing key distinctions from untwisted bilayers, with validation by atomistic Slater–Koster tight-binding (a-SKTB) calculations. 
In Section~\ref{sec:topological_domain_wall_states}, we predict topological domain-wall states at chirality-reversal interfaces ($\zeta=\pm$), mapping the bulk to a Bernevig-Hughes-Zhang-like framework to establish nontrivial topology. Analytical domain wall models reveal a Jackiw-Rebbi-like mechanism tied to chirality reversal, and a-SKTB simulations confirm the robust in-gap modes.
In Section~\ref{sec:discussion_and_conclusion}, we further discuss the resilience of these states under different symmetry-breaking perturbations.

\section{Crystal Structure}
\label{structure}

\begin{figure}
\centering
\includegraphics[width=0.45\textwidth]{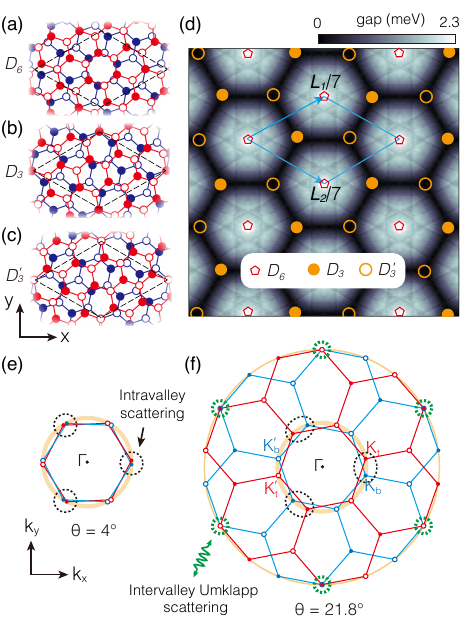} 
\caption{ (a-c) Crystal structures of the $D_6$, $D_3$, and $D_3^\prime$ configurations, respectively. Red/blue colors denote top/bottom layers; solid/hollow markers distinguish $A$/$B$ sublattices. (d) Direct band gap maps of local structures versus in-plane translation vector $\bs{d}$ of top layer. (e) Brillouin zone schematic of monolayer Dirac points at small twist angles: solid/hollow markers denote $K$/$K^{\prime}$ valleys; red/blue colors indicate top/bottom layers. Layer hybridization is dominated by intravalley channel near the Dirac points (dotted black circles). (f) Brillouin zone schematic for $\theta=21.8^{\circ}$ showing $K$/$K^{\prime}$ points of top layer overlapping with $K^{\prime}$/$K$ points of bottom layer at second-zone corners (dotted green circles). Layer hybridization here is dominated by the \textit{intervalley} Umklapp processes.} 
\label{stru}
\end{figure}

Commensurate structures exhibit a well-defined periodicity when the twist angle $\theta$ satisfies~\cite{lopesdossantosGrapheneBilayerTwist2007,renTwistronicsGraphenebasedVan2020} 
\begin{equation}
    \cos\theta=\frac{3m^2+3mn+n^2/2}{3m^2+3mn+n^2},
\end{equation}
where, $m$ and $n$ are coprime positive integers. 
The twist angle $\theta$ is a double-valued function of $(m,n)$, yielding pairs $\pm \theta$ that corresponding to chiral structures with opposite handednesses. 
This structural chirality can be imprinted in electronic structures, driving chiral transport phenomena~\cite{zhaiTimereversalEvenCharge2023a,liDynamicalChiralNernst2024}, which is absent in untwisted bilayers.
The minimal commensurate structure occurs at $(m,n)=(1,1)$ with $\theta\approx \pm21.8^{\circ}$, forming a moir\'{e} superlattice with 28 atoms and primitive vectors $\boldsymbol{L}_{i}$ elongated by $\sqrt{7}$ relative to monolayer graphene~\cite{zhaoUnveilingAtomicScaleMoire2021,liTuningCommensurabilityTwisted2024}.
At this angle, the \textit{intervalley} Umklapp interlayer tunneling is strongest, which is negligible at small twist angles.
Crucially, such small moir\'{e} scale amplifies the non-negligible influence of crystalline symmetry details, which depends sensitively on the rotation center, on low-energy electronic response properties. 
For instance, rotating the top layer around the hexagon center results in commensurate structures belonging to point groups $D_6$ (see Fig~\ref{stru}(a)), referred to here as the $D_{6}$ configuration.
Alternatively, rotating the top layer around an overlapping atom site results in commensurate structures belonging to chiral point groups $D_3$, as shown in Fig~\ref{stru}(b-c). 
This configuration manifests in two distinct types: the $D_{3}$ configuration (see Fig~\ref{stru}(b)) resulting from rotation along the $A$ sublattice in the top layer ($tA$) and the $B$ sublattice in the bottom layer ($bB$), and the $D_{3}^{\prime}$ configuration (see Fig~\ref{stru}(c)), involving rotation along the $B$ sublattice in the top layer ($tB$) and the $A$ sublattice in the bottom layer ($bA$).

One observes that the three configurations in $21.8^{\circ}$ can switch to each other via in-plane translation of the top layer. As shown in Fig~\ref{stru}(d), displacing the top layer by $\boldsymbol{d}=\pm(\boldsymbol{L}_1+\boldsymbol{L}_2)/21$ converts the $D_{6}$ configuration to  $D_{3}$ or $D_{3}^{\prime}$.
Figure~\ref{stru}(d) also maps the direct band gap across various reconstructed structures of tBGs at $\theta=21.8^{\circ}$ as a function of top-layer translation vector $\boldsymbol{d}$. 
Interestingly, the data reveal a real-space moir\'{e} periodicity of $\boldsymbol{L}_i/7$, showing a seven-fold contraction relative to $\boldsymbol{L}_i$~\cite{scheer2022magic,dunbrackIntrinsicallyMultilayerMoire2023}.

In tBGs, space inversion $P$ and all the mirror symmetries are broken. 
At $\theta=21.8^{\circ}$, the $D_{6}$-symmetric configurations possess the highest symmetry (space group No. 177), preserving both the $C_{3z}$ and $C_{2z}$ rotational symmetries inherited from monolayer graphene. While $C_{2x}$ symmetry is retained in both $D_{6}$ and $D_{3}$ configurations, $D_{3}$ (space group No. 150) systems break $C_{2z}$. Beyond these three specific symmetry classes, all in-plane rotational symmetries are broken, leading to reduced symmetry constraints on electronic states. These symmetries will be important for our discussion below.

\section{Effective Low-energy models}
\label{sec:effective_low_energy_models}

Interlayer coupling in tBGs obeys momentum conservation
\begin{equation}
    \boldsymbol{k}+\boldsymbol{G}=\boldsymbol{k}^\prime+\boldsymbol{G}^\prime,
\end{equation}
where $\boldsymbol{k}$ and $\boldsymbol{k}^\prime$ are Bloch wave vectors of the two layers and $\boldsymbol{G}$ and $\boldsymbol{G}^\prime$ are their reciprocal lattice vectors. 
At small twist angles, the intravalley scattering between Dirac cones dominates the low-energy physics of tBG (see Fig.~\ref{stru}(e)), leading to the seminal continuum model of small twist angle tBG~\cite{lopesdossantosContinuumModelTwisted2012,koshinoMaximallyLocalizedWannier2018}.
For larger angles, however, the intravalley coupling hybridizes states far from the Dirac points, while the \textit{intervalley} tunneling governs the low-energy sector~\cite{Umk_PRL_2015,Umk_Nat_2019}.
At $\theta= 21.8^{\circ}$, the \textit{intervalley} Umklapp scattering peaks due to minimized Umklapp scattering wavevector lengths.

\begin{figure}
\centering
\includegraphics[width=0.48\textwidth]{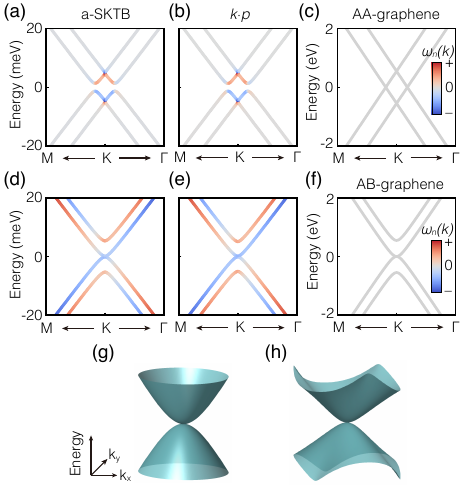} 
\caption{ Comparison of low-energy band structures in large-angle tBG versus untwisted bilayers. (a, d) Bulk bands of the $D_6$ and $D_3$ configurations from atomistic-SKTB method. (b, e) Corresponding $k\cdot p$ effective model results. (c, f) Untwisted $AA$/$AB$-stacked bilayer graphene bands for reference.  Color coding denotes $k$-space  band geometric quantity $\omega_n(\bs{k})$. (g, h) Intra- and interlayer chiral term effects in $D_3$ $k\cdot p$ model: chiral hoppings restore characteristic trigonal warping in low-energy bands (only topmost valence and bottommost conduction bands are shown).}
\label{bands}
\end{figure}

Prior studies~\cite{meleCommensurationInterlayerCoherence2010, moonOpticalAbsorptionTwisted2013} reveal similarities in the four-band low-energy spectra of tBGs at $\theta=21.8^{\circ}$ and untwisted bilayers, where $D_{6}$-, $D_{3}$-, and $D_{3}^{\prime}$-symmetric configurations correspond to $AA$, $AB$, and $BA$ stacking, respectively, as shown in the band structures of Fig.~\ref{bands}(a, c, d, f).
The defining factor for their differing electronic behaviors is the \textit{intervalley} Umklapp tunneling, which should intrinsically encode structural chirality.
To understand its role, we construct effective low-energy $k\cdot p$ Hamiltonians near the moir\'{e} Brillouin zone corners $K$/$K^{\prime}$ under different symmetry constraints, elucidating how this process reshapes electronic structures and band topology.

We start with the $D_{6}$ configuration, the little group at $K$/$K^{\prime}$ point contains the generators $C_{3z}$, $C_{2x}$, which are represented in the low-energy four-state basis at $K$/$K^{\prime}$ point as~\cite{Bradley1972}
\begin{equation}
C_{3z}=-\frac{1}{2}\tau_0\sigma_0+\frac{\sqrt{3}}{2}i\tau_z\sigma_z, ~C_{2x}=-\frac{1}{2}\tau_x\sigma_0-\frac{\sqrt{3}}{2}\tau_y\sigma_z,
\end{equation}
where $\tau_i$ and $\sigma_i$ are the Pauli matrices acting on the effective layer and sublattice index, respectively, $\tau_0$ and $\sigma_0$ are $2\times2$ identity matrix. In addition, we have a combined symmetry $C_{2z}\mathcal{T}=\tau_0\sigma_x\mathcal{K}$, where $\mathcal{T}$ is the time reversal symmetry and $\mathcal{K}$ is the complex conjugation. 
In the basis of this representation, the effective four-band low-energy model of $D_6$ configuration is given by
\begin{equation}
\begin{split}
    &\mathcal{H}_{D_6}(\bs{q})=\left[
    \begin{matrix}
    \hbar v_F\bs{\sigma}\cdot\bs{q}&T\\
    T^\dagger&\hbar v_F\bs{\sigma}\cdot\bs{q}
    \end{matrix}
    \right],\\
    &T=\left[
    \begin{matrix}
        M-i\zeta\lambda&0\\
        0&M+i\zeta\lambda
    \end{matrix}
    \right].
\end{split}
    \label{eq:kpD6}
\end{equation}
The model parameters $M$ (non-chiral interlayer coupling) and $\lambda$ (chiral interlayer coupling) are real-valued, with $\bs{q}$ measured from the $K$/$K^{\prime}$ point, $v_{F}$ the renormalized Fermi velocity, and $\zeta=\pm$ encoding structural handedness.
Symmetry analysis identify $\lambda$ as a hallmark of inversion symmetry breaking, which lifts the degeneracy of mirror-symmetric configurations, splitting the system into chiral left- ($\zeta=+$) or right-handed ($\zeta=-$) ones.
These parameters are determined by fitting low-energy bands near $K$ point, while $\zeta$ is reflected in both the energy gap as well as the band geometric quantity $\omega_{n}(\bs{k})$ (c.f. Eq. (B2) in Apendix B), which is sensitive to structural chirality~\cite{zhaiTimereversalEvenCharge2023a}.
As shown in Fig.~\ref{bands}(a,b), our effective model reproduces both the energy dispersions and band geometric responses of the a-SKTB framework (more details in Appendix A), confirming its fidelity in capturing $D_{6}$-symmetric tBG’s low-energy physics.

As to $D_3$ configuration, it breaks $C_{2z}$ symmetry, the remaining rotational operators are represented as
\begin{equation}
    \begin{split}
        &C_{3z}=(\frac{1}{2}\tau_0+\frac{\sqrt{3}}{2}i\tau_z)(\frac{1}{2}\sigma_0-\frac{\sqrt{3}}{2}i\sigma_z),\\
        &C_{2x}=(\frac{1}{2}\tau_x-\frac{\sqrt{3}}{2}\tau_y)(\frac{1}{2}\sigma_x+\frac{\sqrt{3}}{2}\sigma_y).
    \end{split}
\end{equation}
Thus the effective model up to $q^{2}$ order is
\begin{equation}
\begin{split}
    &\mathcal{H}_{D_3}(\bs{q})=\left[
    \begin{matrix}
    \hbar v_F\bs{\sigma}\cdot\bs{q}+f(\bs{q})&T\\
    T^\dagger&\hbar v_F\bs{\sigma}\cdot\bs{q}-f(\bs{q})
    \end{matrix}
    \right],\\
    &f(\bs{q})=\zeta\frac{\hbar^2 v_{\|} 2}{2m}\bs{\sigma}\cdot(2q_xq_y,~q_x^2-q_y^2),\\
    &T=\left[
    \begin{matrix}
        -i\zeta\hbar v_{\perp}(q_x+iq_y)&M\\
        0&i\zeta\hbar v_{\perp}(q_x+iq_y)
    \end{matrix}
    \right].
\end{split}
    \label{eq:kpD3}
\end{equation}
Unlike the gapped $D_{6}$ phase, the $D_{3}$-symmetric configuration exhibits semimetallic behavior, characterized by a quadratic band crossing at the $K$ point (see Fig.~\ref{bands}(d-e)). 
We stress that the $q$-quadratic terms in model Eq.~\ref{eq:kpD3} are indispensable to capture the correct band geometric quantity and trigonal warping of band structures (see Fig.~\ref{bands}(g, h)).
Furthermore, interlayer chiral coupling in this stacking manifests as $p$-wave term, contrasting with the symmetry-protected constant coupling in the $D_{6}$ model (Eq.~\ref{eq:kpD6}).
Note that the intralayer chirality produces an inequivalence between top and bottom layers, possibly originating from projection of higher-energy band information to lower-energy bands. For the $D_3^\prime$ configuration, since it is linked to the $D_3$ configuration by an $C_{2z}$ operation, they are distinguished by the interlayer coupling given by
\begin{equation}
    T^{\prime}=\left[
    \begin{matrix}
        -i\zeta\hbar v_{\perp}(q_x+iq_y)&0\\
        M&i\zeta\hbar v_{\perp}(q_x+iq_y)
    \end{matrix}
    \right],
\end{equation}
while the intralayer terms remain unchanged.

\section{Topological domain-wall states}
\label{sec:topological_domain_wall_states}

The interlayer terms in the effective model Eq.~\ref{eq:kpD6} partition into $P$-preserving ($M$) and $P$-breaking ($\lambda$) contributions.
Removing the latter reduces the Hamiltonian to the well-known $AA$-stacked bilayer graphene, while retaining it introduces a binary parameter $\zeta=\pm$. 
Although the $\zeta=+$ and $\zeta=-$ configurations exhibit identical band dispersions and second-order band topology~\cite{Lee_tBGHOTI_2019,WangZF_2021_PRL}, the potential for topological protected domain-wall modes at their interface is an open question. To explore this, we construct a hybrid system with a domain wall between $\mathcal{H}_{D_6}$ and $P^{-1} \mathcal{H}_{D_6} P$ and investigate its electronic structure.

\begin{figure}
\centering
\includegraphics[width=0.48\textwidth]{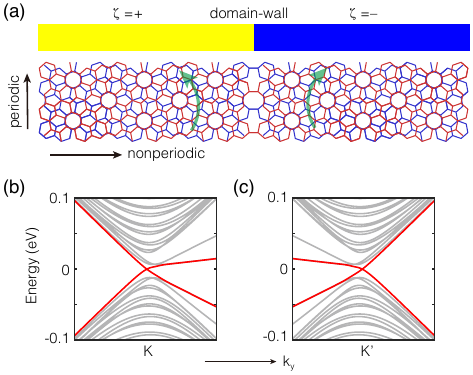} 
\caption{Topological domain-wall states induced by structural chirality. (a) Schematic of a domain wall between $\theta = \pm 21.8^\circ$ tBG ($D_6$ configuration). A short length is shown for  better representation.  (b, c) Corresponding energy spectra.}
\label{DWband}
\end{figure}

To have a quantitative description, we derive a domain-wall model from the bulk $\mathcal{H}_{D_6}$.
First, consider a flat edge at $x=0$, with left-handed tBG ($\zeta=+$) occupying $x<0$ and right-handed tBG ($\zeta=-$) occupying $x>0$ (see Fig.~\ref{DWband}(a)). 
The domain-wall states are solved from the following Schrödinger equation
\begin{equation}
\tilde{\mathcal{H}} \psi=E \psi,
\end{equation}
with
\begin{equation}
 \tilde{\mathcal{H}} = \hbar v_F(-i\partial_x)\tau_0\sigma_x+M\tau_x\sigma_0+\zeta(x)\lambda\tau_y\sigma_z.
    \label{eq:kpD6semi}
\end{equation}
Here $q_{y}$ is a good quantum number, while $q_{x}$ is replaced by $-i\hbar\partial_{x}$.
At $q_{y}=0$, we find two exponentially localized zero-energy modes 
\begin{equation}
\psi_{1,2} \propto \exp \left(\frac{ \pm i M+\zeta \lambda}{\hbar v_F} x\right) \otimes\binom{1}{ \pm i}_\tau \otimes\binom{1}{ \pm i}_\sigma.
\end{equation}
Projecting the $\tilde{\mathcal{H}}$ onto these states yields a 1D edge model 
\begin{equation}
\mathcal{H}_{\text{edge}}=\hbar v_F q_y \sigma_y,
\end{equation}
where $q_y$ is the wave vector along the boundary, and the Pauli
matrices $\sigma$ act on the space of $(\psi_1, \psi_2)^{T}$. 

We demonstrate that topological protected domain-wall states exist at interfaces between systems even with identical bulk bands and topological invariants. 
From the above derivation, one notes taht the band-crossing behavior in edge spectra arises from the sign reversal of an intrinsic bulk parameter ($\lambda$), mirroring the Jackiw-Rebbi mechanism~\cite{jackiwSolitonsFermionNumber1976,shenTopologicalInsulatorsDirac2012}, where mass inversion underpins topological edge states.
When interlayer hopping $M$ vanishes, the eigenstates $\psi_1$ and $\psi_2$ are also eigenstates of $\hbar v_F q_y\tau_0\sigma_y$ with eigenvalues $\pm\hbar v_F$, describing chiral modes counterpropagating along $y$ axis.
Finite $M$ hybridizes these pseudospin channels without destroying intrinsic topological nature, mirroring the effect of an in-plane Zeeman field in a Bernevig-Hughes-Zhang(BHZ) model~\cite{bernevigQuantumSpinHall2006}.
Notably, one could map the Hamiltonian in Eq.~\ref{eq:kpD6} (without $M$) to the well-known BHZ model
\begin{equation}
    U^\dagger\mathcal{H}_{D_6}U=\hbar v_F\tau_0\boldsymbol{\sigma}\cdot\boldsymbol{q}+\zeta\frac{\Delta}{2}\tau_z\sigma_z,
    \label{BHZ}
\end{equation}
which describes the quantum spin Hall effect with opposite spin currents. Finite $M$ breaks time-reversal symmetry but preserves the bulk gap, maintaining its nontrivial topology~\cite{Ren_2020_PRL,cchen_2020_PRL}.

Our analysis reveals that two regions of $D_6$-symmetric tBGs related by spatial inversion $P$ must host topological protected domain-wall states near the $K$/$K^{\prime}$ valley.
To verify this, we construct a tBG ribbon with right-handed tBG on the left side and left-handed tBG on the right side, imposing periodicity along the domain-wall boundary (see Fig.~\ref{DWband}(a)).
Numerical a-SKTB calculations of the energy spectrum confirm the existence of these topological states that manifest themselves as in-gap band-crossing modes (red lines in Fig.~\ref{DWband}(b-c)).

\section{Discussion and conclusion}
\label{sec:discussion_and_conclusion}

\begin{figure*}
\centering
\includegraphics[width=0.8\textwidth]{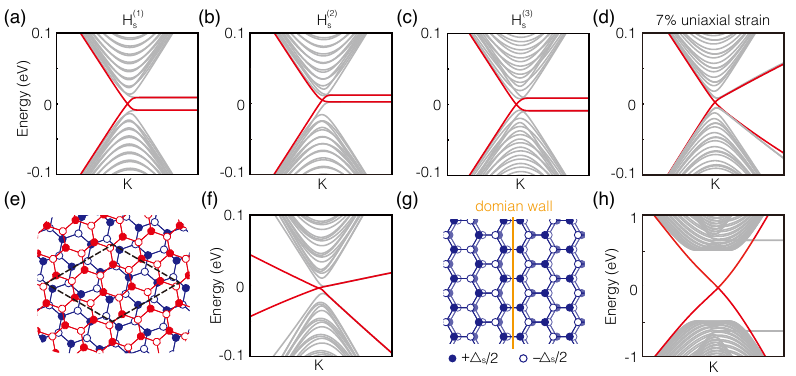} 
\caption{Topological robustness against different symmetry-breaking perturbations. (a-d) Energy spectra for the domain walls under $H_s^{(1)}, H_s^{(2)}, H_s^{(3)}$ ($\Delta_s=15$ meV) and $7\%$ uniaxial strain, respectively. (e, f) Crystal structure and domain-wall states for a configuration displaced from $D_6$ structure by $\bs{L}_1/21$. (g, h) Alternative realization of the topological domain-wall states.  }
\label{fig_perb}
\end{figure*}

Topological domain-wall states are robust against weak symmetry-breaking perturbations as long as the bulk gap remains intact. 
To systematically probe this robustness, we introduce different symmetry-breaking on-site potentials and a uniaxial strain perturbation. Effective tight-binding simulations (more details in Appendix B) demonstrate the persistence of these states under such perturbations.

First, we incorporate a  $C_{2z}$-breaking staggered potential term $H_{s}^{(1)}= (\Delta_s/2)\zeta\tau_z\sigma_z$ into the model (Eq.~\ref{eq:kpD6semi}).
The bulk gap remains open for all $\Delta_s$, and band-crossing behavior is observed (see Fig.~\ref{fig_perb}(a)), confirming the stability of the topological states.
Next, we show that a distinct $C_{2z}$- and $C_{2x}$-breaking term $H_{s}^{(2)}= (\Delta_s/2)\zeta\tau_0\sigma_z$ still keep band-crossing behavior (see Fig.~\ref{fig_perb}(b)).
A third perturbation, $H_{s}^{(3)}= (\Delta_s/2)\tau_z\sigma_0$, which breaks $C_{2x}$ symmetry, also preserves band-crossing behavior (see Fig.~\ref{fig_perb}(c)), further confirming its topological robustness.
Additionally, we apply an unrealistic 7$\%$ uniaxial strain along the $x$-axis in Fig~\ref{DWband}(a),  breaking $C_{3z}$. The bulk gap remains open, leaving the domain-wall states unaffected (see Fig.~\ref{fig_perb}(d)).

The above analysis demonstrates that topological domain-wall states persist under various symmetry-breaking perturbations. This is also verified by a-SKTB calculations (see Appendix A).
Furthermore, we claim that this conclusion extends universally to tBG systems beyond the high-symmetric $D_6$ configuration.
For generic tBG systems with arbitrary interlayer displacements, the effective Hamiltonian can be expressed as $\mathcal{H}_{D_6}+\Delta$, where $\Delta$ represents certain symmetry-breaking deviations induced by displacement.
Crucially, topological domain-wall states should survive provided $\Delta$ preserves the bulk gap, avoiding closure (e.g.,  gapless regimes in Fig.~\ref{stru}(d)).
This universality is exemplified for $\theta=21.8^{\circ}$, where band-crossing behavior survives even in randomly displaced tBGs (see Fig~\ref{fig_perb}(e, f)).
Notably, while $\zeta$-reversed domain-wall states vanish in gapless regimes, valley-Chern-dominated topological channels emerge at $D_3/D_3^{\prime}$ interfaces (see Appendix C), suggesting competing topological mechanisms.

The realization of these topological domain-wall states is constrained by the \textit{intervalley} Umklapp scattering strength, here estimated at $\sim 10$ meV.
Crucially, as derived from Eq.~\ref{BHZ}, when $\sigma_i$ operates on the layer index, our $k\cdot p$ model can also map to bilayer graphene with certain staggered potential. 
This equivalence enables the engineering of the topological domain-wall states at interfaces between two $AA$-stacked bilayers.
Specifically, on the left side of the interface, sublattices $tA$ and $bB$ are assigned with onsite energy $+\Delta/2$, while $tB$ and $bA$ on the right side are set to $-\Delta/2$ (see Fig~\ref{fig_perb}(g)). 
Our calculations confirm the emergence of robust domain-wall states in this configuration (see Fig~\ref{fig_perb}(h)), which indicates that the paradigm extends beyond graphene: analogous states arise at domain walls between  $AA^\prime$-stacking and  $A^\prime A$-stacking bilayers such as hexagonal boron nitride, proving broad applicability in vdW systems.


\begin{appendix}
\label{aSKTB_supp}

\renewcommand{\theequation}{A\arabic{equation}}
\setcounter{equation}{0}
\renewcommand{\thefigure}{A\arabic{figure}}
\setcounter{figure}{0}
\renewcommand{\thetable}{A\arabic{table}}
\setcounter{table}{0}

\section{The atomistic SKTB model of tBG}

The atomistic SKTB model of bilayer graphene is given by~\cite{moonOpticalAbsorptionTwisted2013}
\begin{equation}
    \mathcal{H}=-\sum_{\langle i,j\rangle}t(\bs{d}_{ij})c_i^\dagger c_j+h.c.,
\end{equation}
where $c_i^\dagger$ and $c_j$ denote the creation and annihilation operators for the $p_z$ orbital on site $i$ and $j$, respectively, $\bs{d}_{ij}$ symbolizes the position vector from site $i$ to $j$, and $t(\bs{d}_{ij})$ represents the hopping amplitude between site $i$ and $j$. We adopt the following approximations:
\begin{equation}
\begin{split}
    -&t(\bs{d})=V_{pp\pi}\Bigg[1-\Big(\frac{\bs{d}\cdot\bs{e}_z}{d}\Big)^2\Bigg]+V_{pp\sigma}\Big(\frac{\bs{d}\cdot\bs{e}_z}{d}\Big)^2,\\
    &V_{pp\pi}=V_{pp\pi}^0\exp\Big(-\frac{d-a_0}{\delta_0}\Big),\\
    &V_{pp\sigma}=V_{pp\sigma}^0\exp\Big(-\frac{d-d_0}{\delta_0}\Big).
\end{split}
\end{equation}
In the above, $a_0\approx1.42\;\text{\AA}$ is the nearest-neighbor distance on monolayer graphene, $d_0\approx3.35\;\text{\AA}$ represents the interlayer spacing, $V_{pp\pi}^0$ is the intralayer hopping energy between nearest-neighbor sites, and $V_{pp\sigma}^0$ corresponds to the energy between vertically stacked atoms on the two layers. Here we take $V_{pp\pi}^0\approx-4.32$ eV, $V_{pp\sigma}^0\approx0.78$ eV, and $\delta_0=0.45255\;\text{\AA}$ to fit the dispersions of tBG from DFT result in ref~\cite{shallcrossQuantumInterferenceTwist2008,Chen_2025_ROPP}. Hopping for $d>6\;\text{\AA}$ is exponentially small and thus neglected in our calculation.

\begin{figure}
\centering
\includegraphics[width=0.48\textwidth]{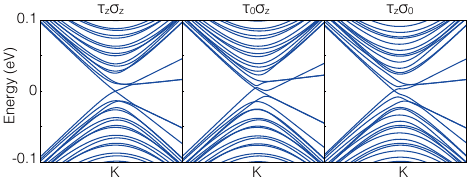} 
\caption{ The atomistic-SKTB band structures of domain walls under three kinds of symmetry-breaking perturbations $(\Delta_s/2)\tau_z\sigma_z$, $(\Delta_s/2)\tau_0\sigma_z$ and $(\Delta_s/2)\tau_z\sigma_0$.}
\label{fig:appendix:sktb}
\end{figure}

To check the topological robustness, we use the a-SKTB model to calculate the domain-wall states under three kinds of symmetry-breaking perturbations $H_s^{(1)}$, $H_s^{(2)}$ and $H_s^{(3)}$ in the maintext. To transform from the effective basis to the atomic basis, the perturbations now become $(\Delta_s/2)\tau_z\sigma_z$, $(\Delta_s/2)\tau_0\sigma_z$ and $(\Delta_s/2)\tau_z\sigma_0$ applied on the sublattices $tA$, $tB$, $bA$ and $bB$ of bilayer graphene, respectively. As Fig.~\ref{fig:appendix:sktb} shows, the survival of the band-crossing behavior verify the robustness of the topological domain-wall states and the effectiveness of our simplified tight-binding model in Appendix B.

\section{Effective simplified tight-binding model}
\renewcommand{\theequation}{B\arabic{equation}}
\setcounter{equation}{0}
\renewcommand{\thefigure}{B\arabic{figure}}
\setcounter{figure}{0}
\renewcommand{\thetable}{B\arabic{table}}
\setcounter{table}{0}

\begin{figure}
\centering
\includegraphics[width=0.45\textwidth]{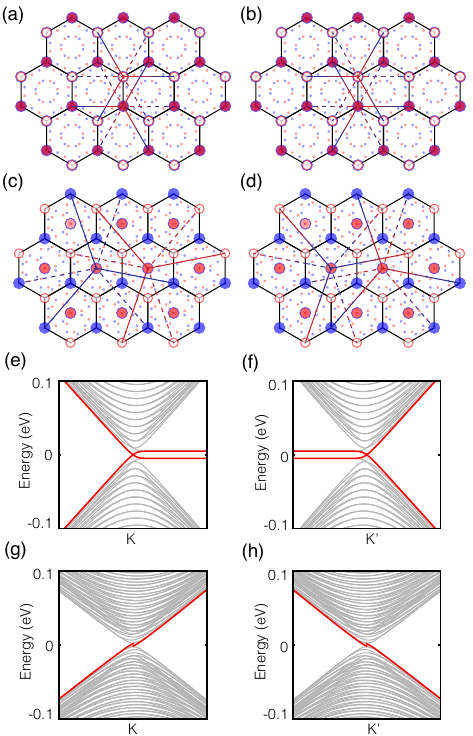} 
\caption{The diagram of chiral hopping in our sim-TB model. The small/large circles represent real/effective atoms. The red and blue colors represent atoms in top and bottom layers, respectively. The empty and solid circles represent sublattice or effective sublattice $A$ and $B$, respectively.  The pure-colored segments depict the intralayer hopping with the hopping of the top/bottom layer colored by red/ blue. The gradient arrows depict the interlayer hopping with the red/blue terminal expressing sublattice in the top/bottom layer. (a) depicts the chiral hopping of $D_6$ sim-TB model with twist angle $\theta=+21.8^\circ$. (b) gives the chiral hopping of $D_6$ sim-TB model of twist angle $\theta=-21.8^\circ$. Clearly, the chiral hoppings in the two $D_6$ configurations are connected by mirror operation. (c, d) give the intralayer and interlayer chiral hoppings of $D_6$ sim-TB model with twist angle $\theta=+21.8^\circ$, respectively. (e, f) show the bands using $D_6(\zeta=+)$/$D_6(\zeta=-)$ sim-TB model in which the topological domain-wall states are plotted by the red lines. (g, h) show the bands using $D_3$/$D_3^{\prime}$ sim-TB model.} 
\label{simTB}
\end{figure}

We also propose an effective simplified tight-binding model (sim-TB) to further understand the role of $\textit{intervalley}$ Umklapp tunneling and the corresponding topological domain-wall states.
The sim-TB model for the $D_6$ configuration reads
\begin{equation}
    \begin{split}
        \mathcal{H}(\bs{k})&=\chi_1(\bs{k})\tau_0\sigma_x+\chi_2(\bs{k})\tau_0\sigma_y+M\tau_x\sigma_0+i\zeta\eta(\bs{k})\tau_y\sigma_z,\\
        &\chi_1+i\chi_2=t_1\sum_{i=1}^3e^{i\bs{k}\cdot\bs{d}_i^{(1)}},\\
        &\eta=2i\lambda\sum_{i=1}^{3}\sin(\bs{k}\cdot\bs{d}_i^{(2)}),
    \end{split}
    \label{eq:SimTBD6}
\end{equation}
where $\sigma_i$ and $\tau_i$ are the Pauli matrices acting on the effective layer and sublattice index, respectively (see Fig.~\ref{simTB}(a,b)).  
 The nearest-neighbor intralayer hopping vectors are given by $\bs{d}_1^{(1)}=\frac{1}{3}\bs{a}_1+\frac{2}{3}\bs{a}_2$, $\bs{d}_2^{(1)}=-\frac{2}{3}\bs{a}_1-\frac{1}{3}\bs{a}_2$ and $\bs{d}_3^{(1)}=\frac{1}{3}\bs{a}_1-\frac{1}{3}\bs{a}_2$ with the amplitude $t_1$. For the interlayer part, $M$ corresponds to the amplitude of direct effective interlayer and intra-sublattice hopping the  while $\lambda$ represents the chiral next-nearest interlayer and inter-sublattice hopping with hopping vectors $\bs{d}_1^{(2)}=\bs{a}_1$, $\bs{d}_2^{(2)}=\bs{a}_2$ and $\bs{d}_3^{(2)}=-\bs{a}_1-\bs{a}_2$ and $\zeta=+$ or $-$.

The model in Eq.~\ref{eq:kpD6} can be obtained from a low-$\bs{q}$ expansion of the sim-TB model (Eq.~\ref{eq:SimTBD6}) at the $K$ valley. 
Thus, the calculation of $E(\bs{k})$ and band geometric quantity $\omega_n(\bs{k})$ is in good agreement with a-SKTB model. Furthermore, when $\lambda=0$, the model reduces to a tight-binding model of $AA$-stacking bilayer graphene whose Bloch basis has well-defined layers and sublattices~\cite{jungTightbindingModelGraphene2013}.

The band geometric quantity $\omega_n(\bs{k})$ is expressed in the form~\cite{zhaiTimereversalEvenCharge2023a}
\begin{equation}
\omega_n(\bs{k})=\hbar\text{Re}\sum_{n_1\neq n}\frac{[\bs{v}_{nn_1}(\bs{k})\times\bs{v}^{\text{sys}}_{n_1n}(\bs{k})]_z}{E_n(\bs{k})-E_{n_1}(\bs{k})},
\end{equation}
where $n$ represents the band index. The term $\bs{v}^{\text{sys}}_{n_1n}(\bs{k})=\langle u_{n_1}(\bs{k})|\frac{1}{2}{\hat\{\bs{v}},\hat{P}^{\text{sys}}\}|u_{n}(\bs{k})\rangle$ involves operator $\hat{P}^{\text{sys}}=(1+\hat{l}^z)/2$, with $\hat{l}^z=\text{diag}(1,-1)$. This operator helps to distinguish between the two enantiomers as it carries information about the layer degree. 

Similarly, we propose a sim-TB model with Wannier-like basis centered on Wyckoff positions of $D_3$ configuration as below:
\begin{equation}
    \begin{split}
        \mathcal{H}(\bs{k})&=\chi_1(\bs{k})\tau_0\sigma_x+\chi_2(\bs{k})\tau_0\sigma_y\\
        &+\eta_1(\bs{k})\tau_x\sigma_z+\eta_2(\bs{k})\tau_y\sigma_z\\
        &+\frac{M}{2}(\tau_x\sigma_x-\tau_y\sigma_y),\\
        \chi_1+i\chi_2&=t\sum_{i=1}^3e^{i\bs{k}\cdot\bs{d}_i^{(1)}}+s\sum_{i=1}^3e^{i\bs{k}\cdot\bs{d}_i^{(2)}}\\
        &+\zeta\lambda_1(\sum_{i=1}^3e^{i\bs{k}\cdot\bs{d}_i^{(3)}}-\sum_{i=1}^3e^{i\bs{k}\cdot\bs{d}_i^{(4)}}),\\
        \eta_1+i\eta_2&=\zeta\lambda_2(\sum_{i=1}^3e^{-i\bs{k}\cdot\bs{d}_i^{(3)}}-\sum_{i=1}^3e^{-i\bs{k}\cdot\bs{d}_i^{(4)}}).
    \end{split}
    \label{eq:SimTBD3}
\end{equation}
The hopping vectors are given by $\bs{d}_1^{(1)}=\frac{1}{3}\bs{a}_1+\frac{2}{3}\bs{a}_2$, $\bs{d}_2^{(1)}=-\frac{2}{3}\bs{a}_1-\frac{1}{3}\bs{a}_2$ and $\bs{d}_3^{(1)}=\frac{1}{3}\bs{a}_1-\frac{1}{3}\bs{a}_2$ with the amplitude $t$,. In this tight-binding model, extra intralayer chiral hoppings $\bs{d}_1^{(3)}=\frac{1}{3}\bs{a}_1-\frac{4}{3}\bs{a}_2$, $\bs{d}_2^{(3)}=-\frac{5}{3}\bs{a}_1-\frac{1}{3}\bs{a}_2$, $\bs{d}_3^{(3)}=\frac{4}{3}\bs{a}_1+\frac{5}{3}\bs{a}_2$, $\bs{d}_1^{(4)}=-\frac{5}{3}\bs{a}_1-\frac{4}{3}\bs{a}_2$, $\bs{d}_2^{(4)}=\frac{4}{3}\bs{a}_1-\frac{1}{3}\bs{a}_2$ and $\bs{d}_3^{(4)}=\frac{1}{3}\bs{a}_1+\frac{5}{3}\bs{a}_2$ with the amplitude $\lambda_1$ and non-chiral second-nearest neighbor hoppings $\bs{d}_1^{(2)}=-\frac{2}{3}\bs{a}_1-\frac{4}{3}\bs{a}_2$, $\bs{d}_2^{(2)}=-\frac{2}{3}\bs{a}_1+\frac{2}{3}\bs{a}_2$ and $\bs{d}_3^{(2)}=\frac{4}{3}\bs{a}_1+\frac{2}{3}\bs{a}_2$ with the amplitude $s$ are introduced. Furthermore, the interlayer chiral hoppings have inverse directions of intralayer chiral hoppings with the amplitude $\lambda_2$. In Fig.~\ref{simTB}(c) and (d), the Wyckoff positions of the space group of the $D_3$ configuration are identical to the atom positions of $AB$-stacking bilayer graphene. 

Our effective sim-TB models reproduce the key topological phenomena: novel topological domain-wall states persist (Fig.~\ref{simTB}(e,f)), and valley Chern domain-wall states emerge (Fig.~\ref{simTB}(g,h)). This demonstrates quantitative agreement with the low-energy physics across all three high-symmetric configurations.


\section{Valley transport in D3/D3$^\prime$ interface}
\renewcommand{\theequation}{C\arabic{equation}}
\setcounter{equation}{0}
\renewcommand{\thefigure}{C\arabic{figure}}
\setcounter{figure}{0}
\renewcommand{\thetable}{C\arabic{table}}
\setcounter{table}{0}

\begin{figure}
\centering
\includegraphics[width=0.45\textwidth]{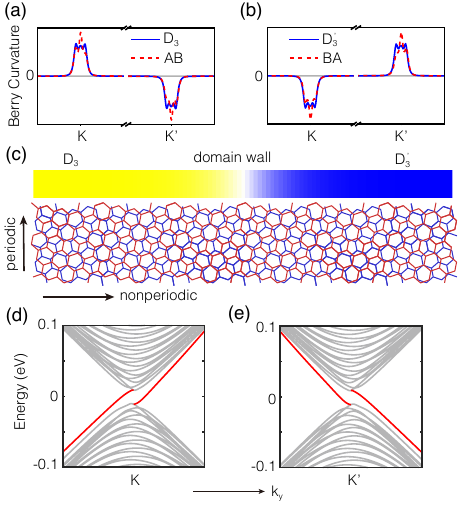} 
\caption{The topological domain-wall states of $D_3$/$D_3^\prime$ interface. (a, b) shows the normalized Berry curvature near valley, showing that valley transport of the $AB$/$BA$ domain wall can also be achieved in the $D_3$/$D_3^\prime$ domain wall. (c) gives the diagram of domain wall between the $D_3$ and $D_3^\prime$ configruation, which is achieved by a slight strain along the nonperiodic direction. (d, e) show the topological domain-wall states of this structure.} 
\label{D3D3prime}
\end{figure}

Figure~\ref{stru}(d) shows the band gap vanishes along the high-symmetry path connecting the $D_3$ and $D_3^\prime$ configurations. This gap closure precludes the chirality-reversal-induced topological domain-wall states discussed in the main text. We demonstrate that valley Chern number protected topological domain-wall states emerge within this gapless regime.

While valley-projected edge modes copropagate at interfaces in gated $AB$/$BA$ bilayer graphene~\cite{zhangValleyChernNumbers2013,vaeziTopologicalEdgeStates2013,juTopologicalValleyTransport2015}, analogous valley currents arise at a $D_3$/$D_3^\prime$ domain wall (see Fig.~\ref{D3D3prime}(c)). This stems from the identical valley Chern numbers of $D_3$ ($D_3^\prime$) and $AB$ ($BA$) stacking bilayer graphene, evidenced by their Berry curvatures (see Fig.~\ref{D3D3prime}(a, b)). Our a-SKTB calculations confirm the valley Chern number protected gapless modes at this interface (see Fig.~\ref{D3D3prime}(d, e)),  which are consistent with Ref.\cite{mondalQuantumValleySubvalley2023}. This result establishes an intrinsic topological valley-conducting channels independent of structural chirality.

\end{appendix}

\bibliographystyle{apsrev4-1}{}
\bibliography{chiral_tBG_ref}

\begin{thebibliography}{47}%
\makeatletter
\providecommand \@ifxundefined [1]{%
 \@ifx{#1\undefined}
}%
\providecommand \@ifnum [1]{%
 \ifnum #1\expandafter \@firstoftwo
 \else \expandafter \@secondoftwo
 \fi
}%
\providecommand \@ifx [1]{%
 \ifx #1\expandafter \@firstoftwo
 \else \expandafter \@secondoftwo
 \fi
}%
\providecommand \natexlab [1]{#1}%
\providecommand \enquote  [1]{``#1''}%
\providecommand \bibnamefont  [1]{#1}%
\providecommand \bibfnamefont [1]{#1}%
\providecommand \citenamefont [1]{#1}%
\providecommand \href@noop [0]{\@secondoftwo}%
\providecommand \href [0]{\begingroup \@sanitize@url \@href}%
\providecommand \@href[1]{\@@startlink{#1}\@@href}%
\providecommand \@@href[1]{\endgroup#1\@@endlink}%
\providecommand \@sanitize@url [0]{\catcode `\\12\catcode `\$12\catcode
  `\&12\catcode `\#12\catcode `\^12\catcode `\_12\catcode `\%12\relax}%
\providecommand \@@startlink[1]{}%
\providecommand \@@endlink[0]{}%
\providecommand \url  [0]{\begingroup\@sanitize@url \@url }%
\providecommand \@url [1]{\endgroup\@href {#1}{\urlprefix }}%
\providecommand \urlprefix  [0]{URL }%
\providecommand \Eprint [0]{\href }%
\providecommand \doibase [0]{http://dx.doi.org/}%
\providecommand \selectlanguage [0]{\@gobble}%
\providecommand \bibinfo  [0]{\@secondoftwo}%
\providecommand \bibfield  [0]{\@secondoftwo}%
\providecommand \translation [1]{[#1]}%
\providecommand \BibitemOpen [0]{}%
\providecommand \bibitemStop [0]{}%
\providecommand \bibitemNoStop [0]{.\EOS\space}%
\providecommand \EOS [0]{\spacefactor3000\relax}%
\providecommand \BibitemShut  [1]{\csname bibitem#1\endcsname}%
\let\auto@bib@innerbib\@empty
\bibitem [{\citenamefont {Andrei}\ and\ \citenamefont
  {MacDonald}(2020)}]{MacD_2020_NatM}%
  \BibitemOpen
  \bibfield  {author} {\bibinfo {author} {\bibfnamefont {E.~Y.}\ \bibnamefont
  {Andrei}}\ and\ \bibinfo {author} {\bibfnamefont {A.~H.}\ \bibnamefont
  {MacDonald}},\ }\href {\doibase 10.1038/s41563-020-00840-0} {\bibfield
  {journal} {\bibinfo  {journal} {Nat. Mater.}\ }\textbf {\bibinfo {volume}
  {19}},\ \bibinfo {pages} {1265} (\bibinfo {year} {2020})}\BibitemShut
  {NoStop}%
\bibitem [{\citenamefont {Kennes}\ \emph {et~al.}(2021)\citenamefont {Kennes},
  \citenamefont {Claassen}, \citenamefont {Xian}, \citenamefont {Georges},
  \citenamefont {Millis}, \citenamefont {Hone}, \citenamefont {Dean},
  \citenamefont {Basov}, \citenamefont {Pasupathy},\ and\ \citenamefont
  {Rubio}}]{Rubio2021moire}%
  \BibitemOpen
  \bibfield  {author} {\bibinfo {author} {\bibfnamefont {D.~M.}\ \bibnamefont
  {Kennes}}, \bibinfo {author} {\bibfnamefont {M.}~\bibnamefont {Claassen}},
  \bibinfo {author} {\bibfnamefont {L.}~\bibnamefont {Xian}}, \bibinfo {author}
  {\bibfnamefont {A.}~\bibnamefont {Georges}}, \bibinfo {author} {\bibfnamefont
  {A.~J.}\ \bibnamefont {Millis}}, \bibinfo {author} {\bibfnamefont
  {J.}~\bibnamefont {Hone}}, \bibinfo {author} {\bibfnamefont {C.~R.}\
  \bibnamefont {Dean}}, \bibinfo {author} {\bibfnamefont {D.}~\bibnamefont
  {Basov}}, \bibinfo {author} {\bibfnamefont {A.~N.}\ \bibnamefont
  {Pasupathy}}, \ and\ \bibinfo {author} {\bibfnamefont {A.}~\bibnamefont
  {Rubio}},\ }\href {\doibase 10.1038/s41567-020-01154-3} {\bibfield  {journal}
  {\bibinfo  {journal} {Nat. Phys.}\ }\textbf {\bibinfo {volume} {17}},\
  \bibinfo {pages} {155} (\bibinfo {year} {2021})}\BibitemShut {NoStop}%
\bibitem [{\citenamefont {Andrei}\ \emph {et~al.}(2021)\citenamefont {Andrei},
  \citenamefont {Efetov}, \citenamefont {Jarillo-Herrero}, \citenamefont
  {MacDonald}, \citenamefont {Mak}, \citenamefont {Senthil}, \citenamefont
  {Tutuc}, \citenamefont {Yazdani},\ and\ \citenamefont
  {Young}}]{Andrea2021marvels}%
  \BibitemOpen
  \bibfield  {author} {\bibinfo {author} {\bibfnamefont {E.~Y.}\ \bibnamefont
  {Andrei}}, \bibinfo {author} {\bibfnamefont {D.~K.}\ \bibnamefont {Efetov}},
  \bibinfo {author} {\bibfnamefont {P.}~\bibnamefont {Jarillo-Herrero}},
  \bibinfo {author} {\bibfnamefont {A.~H.}\ \bibnamefont {MacDonald}}, \bibinfo
  {author} {\bibfnamefont {K.~F.}\ \bibnamefont {Mak}}, \bibinfo {author}
  {\bibfnamefont {T.}~\bibnamefont {Senthil}}, \bibinfo {author} {\bibfnamefont
  {E.}~\bibnamefont {Tutuc}}, \bibinfo {author} {\bibfnamefont
  {A.}~\bibnamefont {Yazdani}}, \ and\ \bibinfo {author} {\bibfnamefont
  {A.~F.}\ \bibnamefont {Young}},\ }\href {\doibase 10.1038/s41578-021-00284-1}
  {\bibfield  {journal} {\bibinfo  {journal} {Nat. Rev. Mater.}\ }\textbf
  {\bibinfo {volume} {6}},\ \bibinfo {pages} {201} (\bibinfo {year}
  {2021})}\BibitemShut {NoStop}%
\bibitem [{\citenamefont {Wilson}\ \emph {et~al.}(2021)\citenamefont {Wilson},
  \citenamefont {Yao}, \citenamefont {Shan},\ and\ \citenamefont
  {Xu}}]{Wilson2021excitons}%
  \BibitemOpen
  \bibfield  {author} {\bibinfo {author} {\bibfnamefont {N.~P.}\ \bibnamefont
  {Wilson}}, \bibinfo {author} {\bibfnamefont {W.}~\bibnamefont {Yao}},
  \bibinfo {author} {\bibfnamefont {J.}~\bibnamefont {Shan}}, \ and\ \bibinfo
  {author} {\bibfnamefont {X.}~\bibnamefont {Xu}},\ }\href {\doibase
  10.1038/s41586-021-03979-1} {\bibfield  {journal} {\bibinfo  {journal}
  {Nature}\ }\textbf {\bibinfo {volume} {599}},\ \bibinfo {pages} {383}
  (\bibinfo {year} {2021})}\BibitemShut {NoStop}%
\bibitem [{\citenamefont {Lau}\ \emph {et~al.}(2022)\citenamefont {Lau},
  \citenamefont {Bockrath}, \citenamefont {Mak},\ and\ \citenamefont
  {Zhang}}]{Zhangfan_2022Nat}%
  \BibitemOpen
  \bibfield  {author} {\bibinfo {author} {\bibfnamefont {C.~N.}\ \bibnamefont
  {Lau}}, \bibinfo {author} {\bibfnamefont {M.~W.}\ \bibnamefont {Bockrath}},
  \bibinfo {author} {\bibfnamefont {K.~F.}\ \bibnamefont {Mak}}, \ and\
  \bibinfo {author} {\bibfnamefont {F.}~\bibnamefont {Zhang}},\ }\href
  {\doibase 10.1038/s41586-021-04173-z} {\bibfield  {journal} {\bibinfo
  {journal} {Nature}\ }\textbf {\bibinfo {volume} {602}},\ \bibinfo {pages}
  {41} (\bibinfo {year} {2022})}\BibitemShut {NoStop}%
\bibitem [{\citenamefont {Carr}\ \emph {et~al.}(2017)\citenamefont {Carr},
  \citenamefont {Massatt}, \citenamefont {Fang}, \citenamefont {Cazeaux},
  \citenamefont {Luskin},\ and\ \citenamefont {Kaxiras}}]{Twist_PRB_2017}%
  \BibitemOpen
  \bibfield  {author} {\bibinfo {author} {\bibfnamefont {S.}~\bibnamefont
  {Carr}}, \bibinfo {author} {\bibfnamefont {D.}~\bibnamefont {Massatt}},
  \bibinfo {author} {\bibfnamefont {S.}~\bibnamefont {Fang}}, \bibinfo {author}
  {\bibfnamefont {P.}~\bibnamefont {Cazeaux}}, \bibinfo {author} {\bibfnamefont
  {M.}~\bibnamefont {Luskin}}, \ and\ \bibinfo {author} {\bibfnamefont
  {E.}~\bibnamefont {Kaxiras}},\ }\href {\doibase 10.1103/PhysRevB.95.075420}
  {\bibfield  {journal} {\bibinfo  {journal} {Phys. Rev. B}\ }\textbf {\bibinfo
  {volume} {95}},\ \bibinfo {pages} {075420} (\bibinfo {year}
  {2017})}\BibitemShut {NoStop}%
\bibitem [{\citenamefont {Hennighausen}\ and\ \citenamefont
  {Kar}(2021)}]{Twistronics_2021}%
  \BibitemOpen
  \bibfield  {author} {\bibinfo {author} {\bibfnamefont {Z.}~\bibnamefont
  {Hennighausen}}\ and\ \bibinfo {author} {\bibfnamefont {S.}~\bibnamefont
  {Kar}},\ }\href {\doibase 10.1088/2516-1075/abd957} {\bibfield  {journal}
  {\bibinfo  {journal} {Electron. Struct.}\ }\textbf {\bibinfo {volume} {3}},\
  \bibinfo {pages} {014004} (\bibinfo {year} {2021})}\BibitemShut {NoStop}%
\bibitem [{\citenamefont {Sun}\ \emph {et~al.}(2024)\citenamefont {Sun},
  \citenamefont {Suriyage}, \citenamefont {Khan}, \citenamefont {Gao},
  \citenamefont {Zhao}, \citenamefont {Liu}, \citenamefont {Hasan},
  \citenamefont {Rahman}, \citenamefont {Chen}, \citenamefont {Lam},\ and\
  \citenamefont {Lu}}]{LuY2024ChemRev}%
  \BibitemOpen
  \bibfield  {author} {\bibinfo {author} {\bibfnamefont {X.}~\bibnamefont
  {Sun}}, \bibinfo {author} {\bibfnamefont {M.}~\bibnamefont {Suriyage}},
  \bibinfo {author} {\bibfnamefont {A.~R.}\ \bibnamefont {Khan}}, \bibinfo
  {author} {\bibfnamefont {M.}~\bibnamefont {Gao}}, \bibinfo {author}
  {\bibfnamefont {J.}~\bibnamefont {Zhao}}, \bibinfo {author} {\bibfnamefont
  {B.}~\bibnamefont {Liu}}, \bibinfo {author} {\bibfnamefont {M.~M.}\
  \bibnamefont {Hasan}}, \bibinfo {author} {\bibfnamefont {S.}~\bibnamefont
  {Rahman}}, \bibinfo {author} {\bibfnamefont {R.-s.}\ \bibnamefont {Chen}},
  \bibinfo {author} {\bibfnamefont {P.~K.}\ \bibnamefont {Lam}}, \ and\
  \bibinfo {author} {\bibfnamefont {Y.}~\bibnamefont {Lu}},\ }\href {\doibase
  10.1021/acs.chemrev.3c00627} {\bibfield  {journal} {\bibinfo  {journal}
  {Chem. Rev.}\ }\textbf {\bibinfo {volume} {124}},\ \bibinfo {pages} {1992}
  (\bibinfo {year} {2024})}\BibitemShut {NoStop}%
\bibitem [{\citenamefont {Cao}\ \emph {et~al.}(2018{\natexlab{a}})\citenamefont
  {Cao}, \citenamefont {Fatemi}, \citenamefont {Fang}, \citenamefont
  {Watanabe}, \citenamefont {Taniguchi}, \citenamefont {Kaxiras},\ and\
  \citenamefont {{Jarillo-Herrero}}}]{CaoY2018_SuperCon}%
  \BibitemOpen
  \bibfield  {author} {\bibinfo {author} {\bibfnamefont {Y.}~\bibnamefont
  {Cao}}, \bibinfo {author} {\bibfnamefont {V.}~\bibnamefont {Fatemi}},
  \bibinfo {author} {\bibfnamefont {S.}~\bibnamefont {Fang}}, \bibinfo {author}
  {\bibfnamefont {K.}~\bibnamefont {Watanabe}}, \bibinfo {author}
  {\bibfnamefont {T.}~\bibnamefont {Taniguchi}}, \bibinfo {author}
  {\bibfnamefont {E.}~\bibnamefont {Kaxiras}}, \ and\ \bibinfo {author}
  {\bibfnamefont {P.}~\bibnamefont {{Jarillo-Herrero}}},\ }\href {\doibase
  10.1038/nature26160} {\bibfield  {journal} {\bibinfo  {journal} {Nature}\
  }\textbf {\bibinfo {volume} {556}},\ \bibinfo {pages} {43} (\bibinfo {year}
  {2018}{\natexlab{a}})}\BibitemShut {NoStop}%
\bibitem [{\citenamefont {Cao}\ \emph {et~al.}(2018{\natexlab{b}})\citenamefont
  {Cao}, \citenamefont {Fatemi}, \citenamefont {Demir}, \citenamefont {Fang},
  \citenamefont {Tomarken}, \citenamefont {Luo}, \citenamefont
  {{Sanchez-Yamagishi}}, \citenamefont {Watanabe}, \citenamefont {Taniguchi},
  \citenamefont {Kaxiras}, \citenamefont {Ashoori},\ and\ \citenamefont
  {{Jarillo-Herrero}}}]{CaoY_2018_Correlated}%
  \BibitemOpen
  \bibfield  {author} {\bibinfo {author} {\bibfnamefont {Y.}~\bibnamefont
  {Cao}}, \bibinfo {author} {\bibfnamefont {V.}~\bibnamefont {Fatemi}},
  \bibinfo {author} {\bibfnamefont {A.}~\bibnamefont {Demir}}, \bibinfo
  {author} {\bibfnamefont {S.}~\bibnamefont {Fang}}, \bibinfo {author}
  {\bibfnamefont {S.~L.}\ \bibnamefont {Tomarken}}, \bibinfo {author}
  {\bibfnamefont {J.~Y.}\ \bibnamefont {Luo}}, \bibinfo {author} {\bibfnamefont
  {J.~D.}\ \bibnamefont {{Sanchez-Yamagishi}}}, \bibinfo {author}
  {\bibfnamefont {K.}~\bibnamefont {Watanabe}}, \bibinfo {author}
  {\bibfnamefont {T.}~\bibnamefont {Taniguchi}}, \bibinfo {author}
  {\bibfnamefont {E.}~\bibnamefont {Kaxiras}}, \bibinfo {author} {\bibfnamefont
  {R.~C.}\ \bibnamefont {Ashoori}}, \ and\ \bibinfo {author} {\bibfnamefont
  {P.}~\bibnamefont {{Jarillo-Herrero}}},\ }\href {\doibase
  10.1038/nature26154} {\bibfield  {journal} {\bibinfo  {journal} {Nature}\
  }\textbf {\bibinfo {volume} {556}},\ \bibinfo {pages} {80} (\bibinfo {year}
  {2018}{\natexlab{b}})}\BibitemShut {NoStop}%
\bibitem [{\citenamefont {Seyler}\ \emph {et~al.}(2019)\citenamefont {Seyler},
  \citenamefont {Rivera}, \citenamefont {Yu}, \citenamefont {Wilson},
  \citenamefont {Ray}, \citenamefont {Mandrus}, \citenamefont {Yan},
  \citenamefont {Yao},\ and\ \citenamefont {Xu}}]{Umk_Nat_WY_2019}%
  \BibitemOpen
  \bibfield  {author} {\bibinfo {author} {\bibfnamefont {K.~L.}\ \bibnamefont
  {Seyler}}, \bibinfo {author} {\bibfnamefont {P.}~\bibnamefont {Rivera}},
  \bibinfo {author} {\bibfnamefont {H.}~\bibnamefont {Yu}}, \bibinfo {author}
  {\bibfnamefont {N.~P.}\ \bibnamefont {Wilson}}, \bibinfo {author}
  {\bibfnamefont {E.~L.}\ \bibnamefont {Ray}}, \bibinfo {author} {\bibfnamefont
  {D.~G.}\ \bibnamefont {Mandrus}}, \bibinfo {author} {\bibfnamefont
  {J.}~\bibnamefont {Yan}}, \bibinfo {author} {\bibfnamefont {W.}~\bibnamefont
  {Yao}}, \ and\ \bibinfo {author} {\bibfnamefont {X.}~\bibnamefont {Xu}},\
  }\href {\doibase 10.1038/s41586-019-0957-1} {\bibfield  {journal} {\bibinfo
  {journal} {Nature}\ }\textbf {\bibinfo {volume} {567}},\ \bibinfo {pages}
  {66} (\bibinfo {year} {2019})}\BibitemShut {NoStop}%
\bibitem [{\citenamefont {Wu}\ \emph {et~al.}(2019)\citenamefont {Wu},
  \citenamefont {Lovorn}, \citenamefont {Tutuc}, \citenamefont {Martin},\ and\
  \citenamefont {MacDonald}}]{WuF_TMD_PRL_2019}%
  \BibitemOpen
  \bibfield  {author} {\bibinfo {author} {\bibfnamefont {F.}~\bibnamefont
  {Wu}}, \bibinfo {author} {\bibfnamefont {T.}~\bibnamefont {Lovorn}}, \bibinfo
  {author} {\bibfnamefont {E.}~\bibnamefont {Tutuc}}, \bibinfo {author}
  {\bibfnamefont {I.}~\bibnamefont {Martin}}, \ and\ \bibinfo {author}
  {\bibfnamefont {A.~H.}\ \bibnamefont {MacDonald}},\ }\href {\doibase
  10.1103/PhysRevLett.122.086402} {\bibfield  {journal} {\bibinfo  {journal}
  {Phys. Rev. Lett.}\ }\textbf {\bibinfo {volume} {122}},\ \bibinfo {pages}
  {086402} (\bibinfo {year} {2019})}\BibitemShut {NoStop}%
\bibitem [{\citenamefont {Wang}\ \emph {et~al.}(2020)\citenamefont {Wang},
  \citenamefont {Shih}, \citenamefont {Ghiotto}, \citenamefont {Xian},
  \citenamefont {Rhodes}, \citenamefont {Tan}, \citenamefont {Claassen},
  \citenamefont {Kennes}, \citenamefont {Bai}, \citenamefont {Kim},
  \citenamefont {Watanabe}, \citenamefont {Taniguchi}, \citenamefont {Zhu},
  \citenamefont {Hone}, \citenamefont {Rubio}, \citenamefont {Pasupathy},\ and\
  \citenamefont {Dean}}]{wangCorrelatedElectronicPhases2020}%
  \BibitemOpen
  \bibfield  {author} {\bibinfo {author} {\bibfnamefont {L.}~\bibnamefont
  {Wang}}, \bibinfo {author} {\bibfnamefont {E.-M.}\ \bibnamefont {Shih}},
  \bibinfo {author} {\bibfnamefont {A.}~\bibnamefont {Ghiotto}}, \bibinfo
  {author} {\bibfnamefont {L.}~\bibnamefont {Xian}}, \bibinfo {author}
  {\bibfnamefont {D.~A.}\ \bibnamefont {Rhodes}}, \bibinfo {author}
  {\bibfnamefont {C.}~\bibnamefont {Tan}}, \bibinfo {author} {\bibfnamefont
  {M.}~\bibnamefont {Claassen}}, \bibinfo {author} {\bibfnamefont {D.~M.}\
  \bibnamefont {Kennes}}, \bibinfo {author} {\bibfnamefont {Y.}~\bibnamefont
  {Bai}}, \bibinfo {author} {\bibfnamefont {B.}~\bibnamefont {Kim}}, \bibinfo
  {author} {\bibfnamefont {K.}~\bibnamefont {Watanabe}}, \bibinfo {author}
  {\bibfnamefont {T.}~\bibnamefont {Taniguchi}}, \bibinfo {author}
  {\bibfnamefont {X.}~\bibnamefont {Zhu}}, \bibinfo {author} {\bibfnamefont
  {J.}~\bibnamefont {Hone}}, \bibinfo {author} {\bibfnamefont {A.}~\bibnamefont
  {Rubio}}, \bibinfo {author} {\bibfnamefont {A.~N.}\ \bibnamefont
  {Pasupathy}}, \ and\ \bibinfo {author} {\bibfnamefont {C.~R.}\ \bibnamefont
  {Dean}},\ }\href {\doibase 10.1038/s41563-020-0708-6} {\bibfield  {journal}
  {\bibinfo  {journal} {Nat. Mater.}\ }\textbf {\bibinfo {volume} {19}},\
  \bibinfo {pages} {861} (\bibinfo {year} {2020})}\BibitemShut {NoStop}%
\bibitem [{\citenamefont {Cai}\ \emph {et~al.}(2023)\citenamefont {Cai},
  \citenamefont {Anderson}, \citenamefont {Wang}, \citenamefont {Zhang},
  \citenamefont {Liu}, \citenamefont {Holtzmann}, \citenamefont {Zhang},
  \citenamefont {Fan}, \citenamefont {Taniguchi}, \citenamefont {Watanabe},
  \citenamefont {Ran}, \citenamefont {Cao}, \citenamefont {Fu}, \citenamefont
  {Xiao}, \citenamefont {Yao},\ and\ \citenamefont
  {Xu}}]{caiSignaturesFractionalQuantum2023}%
  \BibitemOpen
  \bibfield  {author} {\bibinfo {author} {\bibfnamefont {J.}~\bibnamefont
  {Cai}}, \bibinfo {author} {\bibfnamefont {E.}~\bibnamefont {Anderson}},
  \bibinfo {author} {\bibfnamefont {C.}~\bibnamefont {Wang}}, \bibinfo {author}
  {\bibfnamefont {X.}~\bibnamefont {Zhang}}, \bibinfo {author} {\bibfnamefont
  {X.}~\bibnamefont {Liu}}, \bibinfo {author} {\bibfnamefont {W.}~\bibnamefont
  {Holtzmann}}, \bibinfo {author} {\bibfnamefont {Y.}~\bibnamefont {Zhang}},
  \bibinfo {author} {\bibfnamefont {F.}~\bibnamefont {Fan}}, \bibinfo {author}
  {\bibfnamefont {T.}~\bibnamefont {Taniguchi}}, \bibinfo {author}
  {\bibfnamefont {K.}~\bibnamefont {Watanabe}}, \bibinfo {author}
  {\bibfnamefont {Y.}~\bibnamefont {Ran}}, \bibinfo {author} {\bibfnamefont
  {T.}~\bibnamefont {Cao}}, \bibinfo {author} {\bibfnamefont {L.}~\bibnamefont
  {Fu}}, \bibinfo {author} {\bibfnamefont {D.}~\bibnamefont {Xiao}}, \bibinfo
  {author} {\bibfnamefont {W.}~\bibnamefont {Yao}}, \ and\ \bibinfo {author}
  {\bibfnamefont {X.}~\bibnamefont {Xu}},\ }\href {\doibase
  10.1038/s41586-023-06289-w} {\bibfield  {journal} {\bibinfo  {journal}
  {Nature}\ }\textbf {\bibinfo {volume} {622}},\ \bibinfo {pages} {63}
  (\bibinfo {year} {2023})}\BibitemShut {NoStop}%
\bibitem [{\citenamefont {Bistritzer}\ and\ \citenamefont
  {MacDonald}(2011)}]{MacD_PNAS_2011}%
  \BibitemOpen
  \bibfield  {author} {\bibinfo {author} {\bibfnamefont {R.}~\bibnamefont
  {Bistritzer}}\ and\ \bibinfo {author} {\bibfnamefont {A.~H.}\ \bibnamefont
  {MacDonald}},\ }\href {\doibase 10.1073/pnas.1108174108} {\bibfield
  {journal} {\bibinfo  {journal} {Proc. Natl. Acad. Sci.}\ }\textbf {\bibinfo
  {volume} {108}},\ \bibinfo {pages} {12233} (\bibinfo {year}
  {2011})}\BibitemShut {NoStop}%
\bibitem [{\citenamefont {Bistritzer}\ and\ \citenamefont
  {MacDonald}(2010)}]{MacD_2010_PRB}%
  \BibitemOpen
  \bibfield  {author} {\bibinfo {author} {\bibfnamefont {R.}~\bibnamefont
  {Bistritzer}}\ and\ \bibinfo {author} {\bibfnamefont {A.~H.}\ \bibnamefont
  {MacDonald}},\ }\href {\doibase 10.1103/PhysRevB.81.245412} {\bibfield
  {journal} {\bibinfo  {journal} {Phys. Rev. B}\ }\textbf {\bibinfo {volume}
  {81}},\ \bibinfo {pages} {245412} (\bibinfo {year} {2010})}\BibitemShut
  {NoStop}%
\bibitem [{\citenamefont {Yu}\ \emph {et~al.}(2015)\citenamefont {Yu},
  \citenamefont {Wang}, \citenamefont {Tong}, \citenamefont {Xu},\ and\
  \citenamefont {Yao}}]{Umk_PRL_2015}%
  \BibitemOpen
  \bibfield  {author} {\bibinfo {author} {\bibfnamefont {H.}~\bibnamefont
  {Yu}}, \bibinfo {author} {\bibfnamefont {Y.}~\bibnamefont {Wang}}, \bibinfo
  {author} {\bibfnamefont {Q.}~\bibnamefont {Tong}}, \bibinfo {author}
  {\bibfnamefont {X.}~\bibnamefont {Xu}}, \ and\ \bibinfo {author}
  {\bibfnamefont {W.}~\bibnamefont {Yao}},\ }\href {\doibase
  10.1103/PhysRevLett.115.187002} {\bibfield  {journal} {\bibinfo  {journal}
  {Phys. Rev. Lett.}\ }\textbf {\bibinfo {volume} {115}},\ \bibinfo {pages}
  {187002} (\bibinfo {year} {2015})}\BibitemShut {NoStop}%
\bibitem [{\citenamefont {Alexeev}\ \emph {et~al.}(2019)\citenamefont
  {Alexeev}, \citenamefont {Ruiz-Tijerina}, \citenamefont {Danovich},
  \citenamefont {Hamer}, \citenamefont {Terry}, \citenamefont {Nayak},
  \citenamefont {Ahn}, \citenamefont {Pak}, \citenamefont {Lee}, \citenamefont
  {Sohn} \emph {et~al.}}]{Umk_Nat_2019}%
  \BibitemOpen
  \bibfield  {author} {\bibinfo {author} {\bibfnamefont {E.~M.}\ \bibnamefont
  {Alexeev}}, \bibinfo {author} {\bibfnamefont {D.~A.}\ \bibnamefont
  {Ruiz-Tijerina}}, \bibinfo {author} {\bibfnamefont {M.}~\bibnamefont
  {Danovich}}, \bibinfo {author} {\bibfnamefont {M.~J.}\ \bibnamefont {Hamer}},
  \bibinfo {author} {\bibfnamefont {D.~J.}\ \bibnamefont {Terry}}, \bibinfo
  {author} {\bibfnamefont {P.~K.}\ \bibnamefont {Nayak}}, \bibinfo {author}
  {\bibfnamefont {S.}~\bibnamefont {Ahn}}, \bibinfo {author} {\bibfnamefont
  {S.}~\bibnamefont {Pak}}, \bibinfo {author} {\bibfnamefont {J.}~\bibnamefont
  {Lee}}, \bibinfo {author} {\bibfnamefont {J.~I.}\ \bibnamefont {Sohn}},
  \emph {et~al.},\ }\href {\doibase 10.1038/s41586-019-0986-9} {\bibfield
  {journal} {\bibinfo  {journal} {Nature}\ }\textbf {\bibinfo {volume} {567}},\
  \bibinfo {pages} {81} (\bibinfo {year} {2019})}\BibitemShut {NoStop}%
\bibitem [{\citenamefont {Wallbank}\ \emph {et~al.}(2019)\citenamefont
  {Wallbank}, \citenamefont {Krishna~Kumar}, \citenamefont {Holwill},
  \citenamefont {Wang}, \citenamefont {Auton}, \citenamefont {Birkbeck},
  \citenamefont {Mishchenko}, \citenamefont {Ponomarenko}, \citenamefont
  {Watanabe}, \citenamefont {Taniguchi} \emph {et~al.}}]{Umk_NP_2019}%
  \BibitemOpen
  \bibfield  {author} {\bibinfo {author} {\bibfnamefont {J.}~\bibnamefont
  {Wallbank}}, \bibinfo {author} {\bibfnamefont {R.}~\bibnamefont
  {Krishna~Kumar}}, \bibinfo {author} {\bibfnamefont {M.}~\bibnamefont
  {Holwill}}, \bibinfo {author} {\bibfnamefont {Z.}~\bibnamefont {Wang}},
  \bibinfo {author} {\bibfnamefont {G.}~\bibnamefont {Auton}}, \bibinfo
  {author} {\bibfnamefont {J.}~\bibnamefont {Birkbeck}}, \bibinfo {author}
  {\bibfnamefont {A.}~\bibnamefont {Mishchenko}}, \bibinfo {author}
  {\bibfnamefont {L.}~\bibnamefont {Ponomarenko}}, \bibinfo {author}
  {\bibfnamefont {K.}~\bibnamefont {Watanabe}}, \bibinfo {author}
  {\bibfnamefont {T.}~\bibnamefont {Taniguchi}},  \emph {et~al.},\ }\href
  {\doibase 10.1038/s41567-018-0278-6} {\bibfield  {journal} {\bibinfo
  {journal} {Nat. Phys.}\ }\textbf {\bibinfo {volume} {15}},\ \bibinfo {pages}
  {32} (\bibinfo {year} {2019})}\BibitemShut {NoStop}%
\bibitem [{\citenamefont {Park}\ \emph {et~al.}(2019)\citenamefont {Park},
  \citenamefont {Kim}, \citenamefont {Cho},\ and\ \citenamefont
  {Lee}}]{Lee_tBGHOTI_2019}%
  \BibitemOpen
  \bibfield  {author} {\bibinfo {author} {\bibfnamefont {M.~J.}\ \bibnamefont
  {Park}}, \bibinfo {author} {\bibfnamefont {Y.}~\bibnamefont {Kim}}, \bibinfo
  {author} {\bibfnamefont {G.~Y.}\ \bibnamefont {Cho}}, \ and\ \bibinfo
  {author} {\bibfnamefont {S.}~\bibnamefont {Lee}},\ }\href {\doibase
  10.1103/PhysRevLett.123.216803} {\bibfield  {journal} {\bibinfo  {journal}
  {Phys. Rev. Lett.}\ }\textbf {\bibinfo {volume} {123}},\ \bibinfo {pages}
  {216803} (\bibinfo {year} {2019})}\BibitemShut {NoStop}%
\bibitem [{\citenamefont {Liu}\ \emph {et~al.}(2021)\citenamefont {Liu},
  \citenamefont {Xian}, \citenamefont {Mu}, \citenamefont {Zhao}, \citenamefont
  {Liu}, \citenamefont {Rubio},\ and\ \citenamefont {Wang}}]{WangZF_2021_PRL}%
  \BibitemOpen
  \bibfield  {author} {\bibinfo {author} {\bibfnamefont {B.}~\bibnamefont
  {Liu}}, \bibinfo {author} {\bibfnamefont {L.}~\bibnamefont {Xian}}, \bibinfo
  {author} {\bibfnamefont {H.}~\bibnamefont {Mu}}, \bibinfo {author}
  {\bibfnamefont {G.}~\bibnamefont {Zhao}}, \bibinfo {author} {\bibfnamefont
  {Z.}~\bibnamefont {Liu}}, \bibinfo {author} {\bibfnamefont {A.}~\bibnamefont
  {Rubio}}, \ and\ \bibinfo {author} {\bibfnamefont {Z.~F.}\ \bibnamefont
  {Wang}},\ }\href {\doibase 10.1103/PhysRevLett.126.066401} {\bibfield
  {journal} {\bibinfo  {journal} {Phys. Rev. Lett.}\ }\textbf {\bibinfo
  {volume} {126}},\ \bibinfo {pages} {066401} (\bibinfo {year}
  {2021})}\BibitemShut {NoStop}%
\bibitem [{\citenamefont {Qian}\ \emph {et~al.}(2023)\citenamefont {Qian},
  \citenamefont {Li},\ and\ \citenamefont {Liu}}]{CCLiu_2023PRB}%
  \BibitemOpen
  \bibfield  {author} {\bibinfo {author} {\bibfnamefont {S.}~\bibnamefont
  {Qian}}, \bibinfo {author} {\bibfnamefont {Y.}~\bibnamefont {Li}}, \ and\
  \bibinfo {author} {\bibfnamefont {C.-C.}\ \bibnamefont {Liu}},\ }\href
  {\doibase 10.1103/PhysRevB.108.L241406} {\bibfield  {journal} {\bibinfo
  {journal} {Phys. Rev. B}\ }\textbf {\bibinfo {volume} {108}},\ \bibinfo
  {pages} {L241406} (\bibinfo {year} {2023})}\BibitemShut {NoStop}%
\bibitem [{\citenamefont {Chen}\ \emph {et~al.}(2025)\citenamefont {Chen},
  \citenamefont {Zeng},\ and\ \citenamefont {Yao}}]{Chen_2025_ROPP}%
  \BibitemOpen
  \bibfield  {author} {\bibinfo {author} {\bibfnamefont {C.}~\bibnamefont
  {Chen}}, \bibinfo {author} {\bibfnamefont {X.-T.}\ \bibnamefont {Zeng}}, \
  and\ \bibinfo {author} {\bibfnamefont {W.}~\bibnamefont {Yao}},\ }\href
  {\doibase 10.1088/1361-6633/ad9ed8} {\bibfield  {journal} {\bibinfo
  {journal} {Rep. Prog. Phys.}\ }\textbf {\bibinfo {volume} {88}},\ \bibinfo
  {pages} {018001} (\bibinfo {year} {2025})}\BibitemShut {NoStop}%
\bibitem [{\citenamefont {{Lopes~dos~Santos}}\ \emph
  {et~al.}(2007)\citenamefont {{Lopes~dos~Santos}}, \citenamefont {Peres},\
  and\ \citenamefont {Castro~Neto}}]{lopesdossantosGrapheneBilayerTwist2007}%
  \BibitemOpen
  \bibfield  {author} {\bibinfo {author} {\bibfnamefont {J.~M.~B.}\
  \bibnamefont {{Lopes~dos~Santos}}}, \bibinfo {author} {\bibfnamefont
  {N.~M.~R.}\ \bibnamefont {Peres}}, \ and\ \bibinfo {author} {\bibfnamefont
  {A.~H.}\ \bibnamefont {Castro~Neto}},\ }\href {\doibase
  10.1103/PhysRevLett.99.256802} {\bibfield  {journal} {\bibinfo  {journal}
  {Phys. Rev. Lett.}\ }\textbf {\bibinfo {volume} {99}},\ \bibinfo {pages}
  {256802} (\bibinfo {year} {2007})}\BibitemShut {NoStop}%
\bibitem [{\citenamefont {Ren}\ \emph {et~al.}(2020{\natexlab{a}})\citenamefont
  {Ren}, \citenamefont {Zhang}, \citenamefont {Liu},\ and\ \citenamefont
  {He}}]{renTwistronicsGraphenebasedVan2020}%
  \BibitemOpen
  \bibfield  {author} {\bibinfo {author} {\bibfnamefont {Y.-N.}\ \bibnamefont
  {Ren}}, \bibinfo {author} {\bibfnamefont {Y.}~\bibnamefont {Zhang}}, \bibinfo
  {author} {\bibfnamefont {Y.-W.}\ \bibnamefont {Liu}}, \ and\ \bibinfo
  {author} {\bibfnamefont {L.}~\bibnamefont {He}},\ }\href {\doibase
  10.1088/1674-1056/abbbe2} {\bibfield  {journal} {\bibinfo  {journal} {Chi.
  Phys. B}\ }\textbf {\bibinfo {volume} {29}},\ \bibinfo {pages} {117303}
  (\bibinfo {year} {2020}{\natexlab{a}})}\BibitemShut {NoStop}%
\bibitem [{\citenamefont {Zhai}\ \emph {et~al.}(2023)\citenamefont {Zhai},
  \citenamefont {Chen}, \citenamefont {Xiao},\ and\ \citenamefont
  {Yao}}]{zhaiTimereversalEvenCharge2023a}%
  \BibitemOpen
  \bibfield  {author} {\bibinfo {author} {\bibfnamefont {D.}~\bibnamefont
  {Zhai}}, \bibinfo {author} {\bibfnamefont {C.}~\bibnamefont {Chen}}, \bibinfo
  {author} {\bibfnamefont {C.}~\bibnamefont {Xiao}}, \ and\ \bibinfo {author}
  {\bibfnamefont {W.}~\bibnamefont {Yao}},\ }\href {\doibase
  10.1038/s41467-023-37644-0} {\bibfield  {journal} {\bibinfo  {journal} {Nat.
  Commun.}\ }\textbf {\bibinfo {volume} {14}},\ \bibinfo {pages} {1961}
  (\bibinfo {year} {2023})}\BibitemShut {NoStop}%
\bibitem [{\citenamefont {Li}\ \emph {et~al.}(2024{\natexlab{a}})\citenamefont
  {Li}, \citenamefont {Zhai}, \citenamefont {Xiao},\ and\ \citenamefont
  {Yao}}]{liDynamicalChiralNernst2024}%
  \BibitemOpen
  \bibfield  {author} {\bibinfo {author} {\bibfnamefont {J.}~\bibnamefont
  {Li}}, \bibinfo {author} {\bibfnamefont {D.}~\bibnamefont {Zhai}}, \bibinfo
  {author} {\bibfnamefont {C.}~\bibnamefont {Xiao}}, \ and\ \bibinfo {author}
  {\bibfnamefont {W.}~\bibnamefont {Yao}},\ }\href {\doibase
  10.1007/s44214-024-00059-z} {\bibfield  {journal} {\bibinfo  {journal}
  {Quantum Frontiers}\ }\textbf {\bibinfo {volume} {3}},\ \bibinfo {pages} {11}
  (\bibinfo {year} {2024}{\natexlab{a}})}\BibitemShut {NoStop}%
\bibitem [{\citenamefont {Zhao}\ \emph {et~al.}(2021)\citenamefont {Zhao},
  \citenamefont {Qiao}, \citenamefont {Chan}, \citenamefont {Li}, \citenamefont
  {Dan}, \citenamefont {Ning}, \citenamefont {Zhou}, \citenamefont {Quek},
  \citenamefont {Pennycook},\ and\ \citenamefont
  {Loh}}]{zhaoUnveilingAtomicScaleMoire2021}%
  \BibitemOpen
  \bibfield  {author} {\bibinfo {author} {\bibfnamefont {X.}~\bibnamefont
  {Zhao}}, \bibinfo {author} {\bibfnamefont {J.}~\bibnamefont {Qiao}}, \bibinfo
  {author} {\bibfnamefont {S.~M.}\ \bibnamefont {Chan}}, \bibinfo {author}
  {\bibfnamefont {J.}~\bibnamefont {Li}}, \bibinfo {author} {\bibfnamefont
  {J.}~\bibnamefont {Dan}}, \bibinfo {author} {\bibfnamefont {S.}~\bibnamefont
  {Ning}}, \bibinfo {author} {\bibfnamefont {W.}~\bibnamefont {Zhou}}, \bibinfo
  {author} {\bibfnamefont {S.~Y.}\ \bibnamefont {Quek}}, \bibinfo {author}
  {\bibfnamefont {S.~J.}\ \bibnamefont {Pennycook}}, \ and\ \bibinfo {author}
  {\bibfnamefont {K.~P.}\ \bibnamefont {Loh}},\ }\href {\doibase
  10.1021/acs.nanolett.1c00563} {\bibfield  {journal} {\bibinfo  {journal}
  {Nano Letters}\ }\textbf {\bibinfo {volume} {21}},\ \bibinfo {pages} {3262}
  (\bibinfo {year} {2021})}\BibitemShut {NoStop}%
\bibitem [{\citenamefont {Li}\ \emph {et~al.}(2024{\natexlab{b}})\citenamefont
  {Li}, \citenamefont {Zhang}, \citenamefont {Ha}, \citenamefont {Lin},
  \citenamefont {Dong}, \citenamefont {Gao}, \citenamefont {Liu}, \citenamefont
  {Liu}, \citenamefont {Ryu}, \citenamefont {Kim}, \citenamefont {Jozwiak},
  \citenamefont {Bostwick}, \citenamefont {Watanabe}, \citenamefont
  {Taniguchi}, \citenamefont {Kousa}, \citenamefont {Li}, \citenamefont
  {Rotenberg}, \citenamefont {Khalaf}, \citenamefont {Robinson}, \citenamefont
  {Giustino},\ and\ \citenamefont
  {Shih}}]{liTuningCommensurabilityTwisted2024}%
  \BibitemOpen
  \bibfield  {author} {\bibinfo {author} {\bibfnamefont {Y.}~\bibnamefont
  {Li}}, \bibinfo {author} {\bibfnamefont {F.}~\bibnamefont {Zhang}}, \bibinfo
  {author} {\bibfnamefont {V.-A.}\ \bibnamefont {Ha}}, \bibinfo {author}
  {\bibfnamefont {Y.-C.}\ \bibnamefont {Lin}}, \bibinfo {author} {\bibfnamefont
  {C.}~\bibnamefont {Dong}}, \bibinfo {author} {\bibfnamefont {Q.}~\bibnamefont
  {Gao}}, \bibinfo {author} {\bibfnamefont {Z.}~\bibnamefont {Liu}}, \bibinfo
  {author} {\bibfnamefont {X.}~\bibnamefont {Liu}}, \bibinfo {author}
  {\bibfnamefont {S.~H.}\ \bibnamefont {Ryu}}, \bibinfo {author} {\bibfnamefont
  {H.}~\bibnamefont {Kim}}, \bibinfo {author} {\bibfnamefont {C.}~\bibnamefont
  {Jozwiak}}, \bibinfo {author} {\bibfnamefont {A.}~\bibnamefont {Bostwick}},
  \bibinfo {author} {\bibfnamefont {K.}~\bibnamefont {Watanabe}}, \bibinfo
  {author} {\bibfnamefont {T.}~\bibnamefont {Taniguchi}}, \bibinfo {author}
  {\bibfnamefont {B.}~\bibnamefont {Kousa}}, \bibinfo {author} {\bibfnamefont
  {X.}~\bibnamefont {Li}}, \bibinfo {author} {\bibfnamefont {E.}~\bibnamefont
  {Rotenberg}}, \bibinfo {author} {\bibfnamefont {E.}~\bibnamefont {Khalaf}},
  \bibinfo {author} {\bibfnamefont {J.~A.}\ \bibnamefont {Robinson}}, \bibinfo
  {author} {\bibfnamefont {F.}~\bibnamefont {Giustino}}, \ and\ \bibinfo
  {author} {\bibfnamefont {C.-K.}\ \bibnamefont {Shih}},\ }\href {\doibase
  10.1038/s41586-023-06904-w} {\bibfield  {journal} {\bibinfo  {journal}
  {Nature}\ }\textbf {\bibinfo {volume} {625}},\ \bibinfo {pages} {494}
  (\bibinfo {year} {2024}{\natexlab{b}})}\BibitemShut {NoStop}%
\bibitem [{\citenamefont {Scheer}\ \emph {et~al.}(2022)\citenamefont {Scheer},
  \citenamefont {Gu},\ and\ \citenamefont {Lian}}]{scheer2022magic}%
  \BibitemOpen
  \bibfield  {author} {\bibinfo {author} {\bibfnamefont {M.~G.}\ \bibnamefont
  {Scheer}}, \bibinfo {author} {\bibfnamefont {K.}~\bibnamefont {Gu}}, \ and\
  \bibinfo {author} {\bibfnamefont {B.}~\bibnamefont {Lian}},\ }\href {\doibase
  10.1103/PhysRevB.106.115418} {\bibfield  {journal} {\bibinfo  {journal}
  {Phys. Rev. B}\ }\textbf {\bibinfo {volume} {106}},\ \bibinfo {pages}
  {115418} (\bibinfo {year} {2022})}\BibitemShut {NoStop}%
\bibitem [{\citenamefont {Dunbrack}\ and\ \citenamefont
  {Cano}(2023)}]{dunbrackIntrinsicallyMultilayerMoire2023}%
  \BibitemOpen
  \bibfield  {author} {\bibinfo {author} {\bibfnamefont {A.}~\bibnamefont
  {Dunbrack}}\ and\ \bibinfo {author} {\bibfnamefont {J.}~\bibnamefont
  {Cano}},\ }\href {\doibase 10.1103/PhysRevB.107.235425} {\bibfield  {journal}
  {\bibinfo  {journal} {Phys. Rev. B}\ }\textbf {\bibinfo {volume} {107}},\
  \bibinfo {pages} {235425} (\bibinfo {year} {2023})}\BibitemShut {NoStop}%
\bibitem [{\citenamefont {{Lopes dos Santos}}\ \emph
  {et~al.}(2012)\citenamefont {{Lopes dos Santos}}, \citenamefont {Peres},\
  and\ \citenamefont {Castro~Neto}}]{lopesdossantosContinuumModelTwisted2012}%
  \BibitemOpen
  \bibfield  {author} {\bibinfo {author} {\bibfnamefont {J.~M.~B.}\
  \bibnamefont {{Lopes dos Santos}}}, \bibinfo {author} {\bibfnamefont
  {N.~M.~R.}\ \bibnamefont {Peres}}, \ and\ \bibinfo {author} {\bibfnamefont
  {A.~H.}\ \bibnamefont {Castro~Neto}},\ }\href {\doibase
  10.1103/PhysRevB.86.155449} {\bibfield  {journal} {\bibinfo  {journal} {Phys.
  Rev. B}\ }\textbf {\bibinfo {volume} {86}},\ \bibinfo {pages} {155449}
  (\bibinfo {year} {2012})}\BibitemShut {NoStop}%
\bibitem [{\citenamefont {Koshino}\ \emph {et~al.}(2018)\citenamefont
  {Koshino}, \citenamefont {Yuan}, \citenamefont {Koretsune}, \citenamefont
  {Ochi}, \citenamefont {Kuroki},\ and\ \citenamefont
  {Fu}}]{koshinoMaximallyLocalizedWannier2018}%
  \BibitemOpen
  \bibfield  {author} {\bibinfo {author} {\bibfnamefont {M.}~\bibnamefont
  {Koshino}}, \bibinfo {author} {\bibfnamefont {N.~F.~Q.}\ \bibnamefont
  {Yuan}}, \bibinfo {author} {\bibfnamefont {T.}~\bibnamefont {Koretsune}},
  \bibinfo {author} {\bibfnamefont {M.}~\bibnamefont {Ochi}}, \bibinfo {author}
  {\bibfnamefont {K.}~\bibnamefont {Kuroki}}, \ and\ \bibinfo {author}
  {\bibfnamefont {L.}~\bibnamefont {Fu}},\ }\href {\doibase
  10.1103/PhysRevX.8.031087} {\bibfield  {journal} {\bibinfo  {journal} {Phys.
  Rev. X}\ }\textbf {\bibinfo {volume} {8}},\ \bibinfo {pages} {031087}
  (\bibinfo {year} {2018})}\BibitemShut {NoStop}%
\bibitem [{\citenamefont
  {Mele}(2010)}]{meleCommensurationInterlayerCoherence2010}%
  \BibitemOpen
  \bibfield  {author} {\bibinfo {author} {\bibfnamefont {E.~J.}\ \bibnamefont
  {Mele}},\ }\href {\doibase 10.1103/PhysRevB.81.161405} {\bibfield  {journal}
  {\bibinfo  {journal} {Phys. Rev. B}\ }\textbf {\bibinfo {volume} {81}},\
  \bibinfo {pages} {161405} (\bibinfo {year} {2010})}\BibitemShut {NoStop}%
\bibitem [{\citenamefont {Moon}\ and\ \citenamefont
  {Koshino}(2013)}]{moonOpticalAbsorptionTwisted2013}%
  \BibitemOpen
  \bibfield  {author} {\bibinfo {author} {\bibfnamefont {P.}~\bibnamefont
  {Moon}}\ and\ \bibinfo {author} {\bibfnamefont {M.}~\bibnamefont {Koshino}},\
  }\href {\doibase 10.1103/PhysRevB.87.205404} {\bibfield  {journal} {\bibinfo
  {journal} {Phys. Rev. B}\ }\textbf {\bibinfo {volume} {87}},\ \bibinfo
  {pages} {205404} (\bibinfo {year} {2013})}\BibitemShut {NoStop}%
\bibitem [{\citenamefont {Bradley}\ and\ \citenamefont
  {Cracknell}(1972)}]{Bradley1972}%
  \BibitemOpen
  \bibfield  {author} {\bibinfo {author} {\bibfnamefont {C.~J.}\ \bibnamefont
  {Bradley}}\ and\ \bibinfo {author} {\bibfnamefont {A.~P.}\ \bibnamefont
  {Cracknell}},\ }\href@noop {} {\emph {\bibinfo {title} {The Mathematical
  Theory of Symmetry in Solids}}}\ (\bibinfo  {publisher} {Clarendon},\
  \bibinfo {address} {Oxford},\ \bibinfo {year} {1972})\BibitemShut {NoStop}%
\bibitem [{\citenamefont {Jackiw}\ and\ \citenamefont
  {Rebbi}(1976)}]{jackiwSolitonsFermionNumber1976}%
  \BibitemOpen
  \bibfield  {author} {\bibinfo {author} {\bibfnamefont {R.}~\bibnamefont
  {Jackiw}}\ and\ \bibinfo {author} {\bibfnamefont {C.}~\bibnamefont {Rebbi}},\
  }\href {\doibase 10.1103/PhysRevD.13.3398} {\bibfield  {journal} {\bibinfo
  {journal} {Phys. Rev. D}\ }\textbf {\bibinfo {volume} {13}},\ \bibinfo
  {pages} {3398} (\bibinfo {year} {1976})}\BibitemShut {NoStop}%
\bibitem [{\citenamefont {Shen}(2012)}]{shenTopologicalInsulatorsDirac2012}%
  \BibitemOpen
  \bibfield  {author} {\bibinfo {author} {\bibfnamefont {S.-Q.}\ \bibnamefont
  {Shen}},\ }\href@noop {} {\emph {\bibinfo {title} {Topological
  Insulators}}},\ \bibinfo {series} {Springer Series in Solid-State Sciences},
  Vol.\ \bibinfo {volume} {174}\ (\bibinfo  {publisher} {Springer},\ \bibinfo
  {address} {Berlin, Heidelberg},\ \bibinfo {year} {2012})\BibitemShut
  {NoStop}%
\bibitem [{\citenamefont {Bernevig}\ \emph {et~al.}(2006)\citenamefont
  {Bernevig}, \citenamefont {Hughes},\ and\ \citenamefont
  {Zhang}}]{bernevigQuantumSpinHall2006}%
  \BibitemOpen
  \bibfield  {author} {\bibinfo {author} {\bibfnamefont {B.~A.}\ \bibnamefont
  {Bernevig}}, \bibinfo {author} {\bibfnamefont {T.~L.}\ \bibnamefont
  {Hughes}}, \ and\ \bibinfo {author} {\bibfnamefont {S.-C.}\ \bibnamefont
  {Zhang}},\ }\href {\doibase 10.1126/science.1133734} {\bibfield  {journal}
  {\bibinfo  {journal} {Science}\ }\textbf {\bibinfo {volume} {314}},\ \bibinfo
  {pages} {1757} (\bibinfo {year} {2006})}\BibitemShut {NoStop}%
\bibitem [{\citenamefont {Ren}\ \emph {et~al.}(2020{\natexlab{b}})\citenamefont
  {Ren}, \citenamefont {Qiao},\ and\ \citenamefont {Niu}}]{Ren_2020_PRL}%
  \BibitemOpen
  \bibfield  {author} {\bibinfo {author} {\bibfnamefont {Y.}~\bibnamefont
  {Ren}}, \bibinfo {author} {\bibfnamefont {Z.}~\bibnamefont {Qiao}}, \ and\
  \bibinfo {author} {\bibfnamefont {Q.}~\bibnamefont {Niu}},\ }\href@noop {}
  {\bibfield  {journal} {\bibinfo  {journal} {Phys. Rev. Lett.}\ }\textbf
  {\bibinfo {volume} {124}},\ \bibinfo {pages} {166804} (\bibinfo {year}
  {2020}{\natexlab{b}})}\BibitemShut {NoStop}%
\bibitem [{\citenamefont {Chen}\ \emph {et~al.}(2020)\citenamefont {Chen},
  \citenamefont {Song}, \citenamefont {Zhao}, \citenamefont {Chen},
  \citenamefont {Yu}, \citenamefont {Sheng},\ and\ \citenamefont
  {Yang}}]{cchen_2020_PRL}%
  \BibitemOpen
  \bibfield  {author} {\bibinfo {author} {\bibfnamefont {C.}~\bibnamefont
  {Chen}}, \bibinfo {author} {\bibfnamefont {Z.}~\bibnamefont {Song}}, \bibinfo
  {author} {\bibfnamefont {J.-Z.}\ \bibnamefont {Zhao}}, \bibinfo {author}
  {\bibfnamefont {Z.}~\bibnamefont {Chen}}, \bibinfo {author} {\bibfnamefont
  {Z.-M.}\ \bibnamefont {Yu}}, \bibinfo {author} {\bibfnamefont {X.-L.}\
  \bibnamefont {Sheng}}, \ and\ \bibinfo {author} {\bibfnamefont {S.~A.}\
  \bibnamefont {Yang}},\ }\href@noop {} {\bibfield  {journal} {\bibinfo
  {journal} {Phys. Rev. Lett.}\ }\textbf {\bibinfo {volume} {125}},\ \bibinfo
  {pages} {056402} (\bibinfo {year} {2020})}\BibitemShut {NoStop}%
\bibitem [{\citenamefont {Shallcross}\ \emph {et~al.}(2008)\citenamefont
  {Shallcross}, \citenamefont {Sharma},\ and\ \citenamefont
  {Pankratov}}]{shallcrossQuantumInterferenceTwist2008}%
  \BibitemOpen
  \bibfield  {author} {\bibinfo {author} {\bibfnamefont {S.}~\bibnamefont
  {Shallcross}}, \bibinfo {author} {\bibfnamefont {S.}~\bibnamefont {Sharma}},
  \ and\ \bibinfo {author} {\bibfnamefont {O.~A.}\ \bibnamefont {Pankratov}},\
  }\href {\doibase 10.1103/PhysRevLett.101.056803} {\bibfield  {journal}
  {\bibinfo  {journal} {Phys. Rev. Lett.}\ }\textbf {\bibinfo {volume} {101}},\
  \bibinfo {pages} {056803} (\bibinfo {year} {2008})}\BibitemShut {NoStop}%
\bibitem [{\citenamefont {Jung}\ and\ \citenamefont
  {MacDonald}(2013)}]{jungTightbindingModelGraphene2013}%
  \BibitemOpen
  \bibfield  {author} {\bibinfo {author} {\bibfnamefont {J.}~\bibnamefont
  {Jung}}\ and\ \bibinfo {author} {\bibfnamefont {A.~H.}\ \bibnamefont
  {MacDonald}},\ }\href {\doibase 10.1103/PhysRevB.87.195450} {\bibfield
  {journal} {\bibinfo  {journal} {Phys. Rev. B}\ }\textbf {\bibinfo {volume}
  {87}},\ \bibinfo {pages} {195450} (\bibinfo {year} {2013})}\BibitemShut
  {NoStop}%
\bibitem [{\citenamefont {Zhang}\ \emph {et~al.}(2013)\citenamefont {Zhang},
  \citenamefont {MacDonald},\ and\ \citenamefont
  {Mele}}]{zhangValleyChernNumbers2013}%
  \BibitemOpen
  \bibfield  {author} {\bibinfo {author} {\bibfnamefont {F.}~\bibnamefont
  {Zhang}}, \bibinfo {author} {\bibfnamefont {A.~H.}\ \bibnamefont
  {MacDonald}}, \ and\ \bibinfo {author} {\bibfnamefont {E.~J.}\ \bibnamefont
  {Mele}},\ }\href {\doibase 10.1073/pnas.1308853110} {\bibfield  {journal}
  {\bibinfo  {journal} {Proc. Natl. Acad. Sci.}\ }\textbf {\bibinfo {volume}
  {110}},\ \bibinfo {pages} {10546} (\bibinfo {year} {2013})}\BibitemShut
  {NoStop}%
\bibitem [{\citenamefont {Vaezi}\ \emph {et~al.}(2013)\citenamefont {Vaezi},
  \citenamefont {Liang}, \citenamefont {Ngai}, \citenamefont {Yang},\ and\
  \citenamefont {Kim}}]{vaeziTopologicalEdgeStates2013}%
  \BibitemOpen
  \bibfield  {author} {\bibinfo {author} {\bibfnamefont {A.}~\bibnamefont
  {Vaezi}}, \bibinfo {author} {\bibfnamefont {Y.}~\bibnamefont {Liang}},
  \bibinfo {author} {\bibfnamefont {D.~H.}\ \bibnamefont {Ngai}}, \bibinfo
  {author} {\bibfnamefont {L.}~\bibnamefont {Yang}}, \ and\ \bibinfo {author}
  {\bibfnamefont {E.-A.}\ \bibnamefont {Kim}},\ }\href {\doibase
  10.1103/PhysRevX.3.021018} {\bibfield  {journal} {\bibinfo  {journal} {Phys.
  Rev. X}\ }\textbf {\bibinfo {volume} {3}},\ \bibinfo {pages} {021018}
  (\bibinfo {year} {2013})}\BibitemShut {NoStop}%
\bibitem [{\citenamefont {Ju}\ \emph {et~al.}(2015)\citenamefont {Ju},
  \citenamefont {Shi}, \citenamefont {Nair}, \citenamefont {Lv}, \citenamefont
  {Jin}, \citenamefont {Velasco}, \citenamefont {{Ojeda-Aristizabal}},
  \citenamefont {Bechtel}, \citenamefont {Martin}, \citenamefont {Zettl},
  \citenamefont {Analytis},\ and\ \citenamefont
  {Wang}}]{juTopologicalValleyTransport2015}%
  \BibitemOpen
  \bibfield  {author} {\bibinfo {author} {\bibfnamefont {L.}~\bibnamefont
  {Ju}}, \bibinfo {author} {\bibfnamefont {Z.}~\bibnamefont {Shi}}, \bibinfo
  {author} {\bibfnamefont {N.}~\bibnamefont {Nair}}, \bibinfo {author}
  {\bibfnamefont {Y.}~\bibnamefont {Lv}}, \bibinfo {author} {\bibfnamefont
  {C.}~\bibnamefont {Jin}}, \bibinfo {author} {\bibfnamefont {J.}~\bibnamefont
  {Velasco}}, \bibinfo {author} {\bibfnamefont {C.}~\bibnamefont
  {{Ojeda-Aristizabal}}}, \bibinfo {author} {\bibfnamefont {H.~A.}\
  \bibnamefont {Bechtel}}, \bibinfo {author} {\bibfnamefont {M.~C.}\
  \bibnamefont {Martin}}, \bibinfo {author} {\bibfnamefont {A.}~\bibnamefont
  {Zettl}}, \bibinfo {author} {\bibfnamefont {J.}~\bibnamefont {Analytis}}, \
  and\ \bibinfo {author} {\bibfnamefont {F.}~\bibnamefont {Wang}},\ }\href
  {\doibase 10.1038/nature14364} {\bibfield  {journal} {\bibinfo  {journal}
  {Nature}\ }\textbf {\bibinfo {volume} {520}},\ \bibinfo {pages} {650}
  (\bibinfo {year} {2015})}\BibitemShut {NoStop}%
\bibitem [{\citenamefont {Mondal}\ \emph {et~al.}(2023)\citenamefont {Mondal},
  \citenamefont {Ghadimi},\ and\ \citenamefont
  {Yang}}]{mondalQuantumValleySubvalley2023}%
  \BibitemOpen
  \bibfield  {author} {\bibinfo {author} {\bibfnamefont {C.}~\bibnamefont
  {Mondal}}, \bibinfo {author} {\bibfnamefont {R.}~\bibnamefont {Ghadimi}}, \
  and\ \bibinfo {author} {\bibfnamefont {B.-J.}\ \bibnamefont {Yang}},\ }\href
  {\doibase 10.1103/PhysRevB.108.L121405} {\bibfield  {journal} {\bibinfo
  {journal} {Phys. Rev. B}\ }\textbf {\bibinfo {volume} {108}},\ \bibinfo
  {pages} {L121405} (\bibinfo {year} {2023})}\BibitemShut {NoStop}%
\end{thebibliography}%

\end{document}